\documentclass[prx,aps,twocolumn]{revtex4-1}

\usepackage{times}
\usepackage{graphicx}
\usepackage{float}
\usepackage{latexsym,amsmath,amssymb,bm,euscript}
\usepackage{color}
\usepackage{subfigure}
\usepackage{epstopdf}
\usepackage[colorlinks=true,linkcolor=blue,citecolor=blue]{hyperref}
\usepackage{type1cm}

\usepackage{ulem}
\usepackage{cancel}

\usepackage{appendix}

\begin{document}
\title{Microscopic Diagnosis of Universal Geometric Responses in Fractional Quantum Hall Liquids}

\author{Liangdong Hu$^{1}$, Zhao Liu$^{2}$, and D. N. Sheng$^3$, F. D. M. Haldane$^4$, W. Zhu$^1$}
\affiliation{$^1$ Westlake Institute of Advanced Study,  Westlake University, Hangzhou, 310024, China}
\affiliation{$^2$Zhejiang Institute of Modern Physics, Zhejiang University, Hangzhou 310027, China}
\affiliation{$^3$Department of Physics and Astronomy, California State University, Northridge, CA 91330, USA}
\affiliation{$^4$Department of Physics, Princeton University, Princeton, New Jersey 08544, USA}

\begin{abstract}
Topological quantum liquids contain internal degrees of freedom that
are coupled to geometric response. 
Yet, an explicit and microscopic identification of geometric response remains difficult.   
Here, taking notable fractional quantum Hall (FQH) states as typical examples, we systematically investigate a promising protocol -- the Dehn twist deformation on the torus geometry, 
to probe the geometric response of correlated topological states and establish the relation between such response and the universal properties of pertinent states. 
Based on analytical derivations and numerical simulations, we find that the geometry-induced Berry phase encodes novel features 
for a broad class of FQH states at the Laughlin, hierarchy, Halperin and non-Abelian Moore-Read fillings.
Our findings conclusively demonstrate that the adiabatic Dehn twist deformation 
can faithfully capture the geometry of elementary FQH droplets
and intrinsic modular information including topological spin and chiral central charge.
Our approach provides a powerful way to reveal topological orders of generic FQH states
and allows us to address previously open questions.
\end{abstract}

\date{\today}


\maketitle


\section{Introduction}

Topological phases of matter~\cite{Wen_book} possess a variety of properties which are robust against external perturbations as long as the topology of space where the system is defined is not altered. As a celebrated example, the fractional quantum Hall (FQH) effect~\cite{Tsui1982} formed by two-dimensional interacting particles in strong magnetic fields has attracted broad interest in the past decades. The topologically invariant properties of FQH states, including the quantized electrical~\cite{Laughlin1981,Thouless1982,Niu1985} and thermal Hall conductances~\cite{Bid2010,Banerjee2017,Banerjee2018}, topological ground-state degeneracies \cite{Niu1990}, exotic anyonic quasiparticles~\cite{Laughlin1983,Arovas1984,Moore1991}, and entanglement characteristics~\cite{Levin2006,Preskill2006,Haldane2008}, have been extensively studied from both theoretical and experimental sides.

Despite FQH states are often characterized by their topologically invariant features, these states do have intriguing response to variations of the ambient geometry even if these variations preserve the underlying topology. Two representative examples are the intrinsic ``orbital spin" 
~\cite{Wen1992,Haldane2009,Haldane2011} and the Hall viscosity~\cite{Avron1995,Read2009,Haldane2009}. 
The former can be related to intrinsic metrics which describe deformations of a FQH droplet due to anisotropies in the background space (for instance those induced by tilted or spatially inhomogeneous magnetic fields \cite{Papic2012,Papic2016}), while the latter determines an accumulated Berry phase caused by strains applied to the FQH droplet \cite{Avron1995,Read2009}. %
As the geometric response is closely related to the internal topological structure of FQH liquids, 
it is potentially a powerful diagnosis of the underlying FQH topological order.

Nevertheless, so far most studies about the geometric response heavily relied on effective field theories~\cite{Wen1992,Read2009,Hoyos2012,Saremi2012,DTSon2013,Hughes2013,DTSon2014,Cho2014,Read2015,Wiegmann2015,Gromov2015,Stern2019}
and model wave functions~\cite{Read2011,Fremling2014,YeJePark2014,YZYou2015,Moradi2015}.
Despite of a few attempts to connect it with entanglement contents~\cite{Zaletel2013,HHTu2013},
it remains challenging to directly investigate the evolution and response of wave functions at the microscopic level, especially for generic FQH states without prior knowledge. 

In this work, we aim to fill in this blank by establishing a relation between the geometric response of FQH states and their universal (topological and geometric) properties in the presence of Dehn twist on the torus. This relation is analytically derived under a gauge-fixing scheme instead of relying on physical arguments, and can be readily confirmed by numerical simulations in microscopic models. 
Our main finding is, for a robust FQH phase with a set of degenerate ground states evolving adiabatically during the Dehn twist,
there is an accumulated Berry phase in each ground state which contains both topological and geometric information: 
the Hall viscosity related to the averaged guiding-center spin, 
the sector-dependent topological spin, and the chiral central charge.
These information fully characterizes the underlying topological order.
By using extensive exact diagonalization to track the evolution of ground-state wave functions and calculate the accumulated Berry phase,
we demonstrate the validity of this relation for various FQH states 
at the fermionic and bosonic Laughlin, hierarchy, Halperin and non-Abelian Moore-Read fillings, 
and successfully extract the topological and geometric properties of both model wave functions and Coulomb ground states.
As a byproduct, 
we find that the flow of energy spectra under geometric deformation 
plays as a ``smoking-gun" feature to justify the robustness of FQH liquids.
In this context, we demonstrate that the ground-state degeneracy at $\nu=5/2$ under particle-hole symmetric interactions
is fragile, which challenges the identification of (anti-)Pfaffian state based on finite-size calculations.

\section{Geometric Berry Phase from Dehn Twist} 
\label{geometry}

\begin{figure}[t]
	\includegraphics[width=0.48\textwidth]{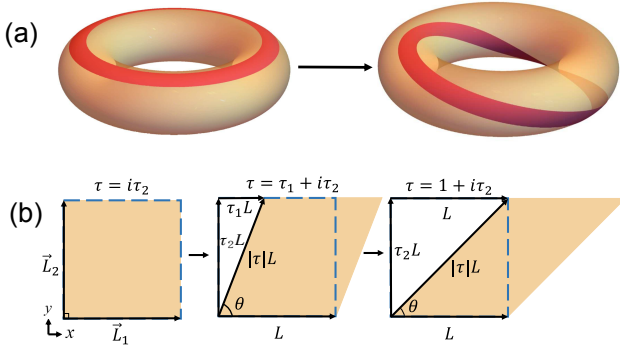}
	\caption{\textbf{Dehn twist operation on the torus.} 
		(a) A twist operation on the annulus (red) illustrates the self-homeomorphism of $\mathcal{T}$ transformation.
		(b) The torus geometry is defined by two fundamental vectors $\vec L_2=L\vec \tau$ and $\vec L_1=L \vec e_x$, 
		and the twist angle is $\theta$. 
		The Dehn twist, i.e., the $\mathcal{T}$ transformation sends $\vec \tau=\tau_1 \vec e_x+\tau_2 \vec e_y$ to its equivalent geometry $\vec \tau+\vec e_x$,
		thus leaves the torus geometry unchanged. The area of the torus does not change during the Dehn twist. Here we give an example of the Dehn twist changing $\vec \tau$ from $\tau_2\vec e_y$ to $\vec e_x+\tau_2\vec e_y$, with $\tau_2=|\vec L_2(\theta=\pi/2)|/|\vec L_1|$. 
	}\label{fig:torus}
\end{figure}

We consider $N_p$ particles with charge $e$ moving in two spatial dimensions on the torus geometry subjected to a perpendicular uniform magnetic field. 
The torus is spanned by two vectors
$\vec L_1 = L \vec e_x$ and $\vec L_2 = L\vec\tau$~\cite{note6},
where $\vec\tau$ is parametrized by the twist angle $\theta$ as
$\vec \tau= \tau_1\vec e_x+\tau_2 \vec e_y=(\cos\theta \vec e_x+ \sin\theta \vec e_y)|\vec L_2|/|\vec L_1|$ such that $\vec L_2=\vec L_2(\theta)$ depends on $\theta$ (Fig.~\ref{fig:torus}). Here $\vec e_x$ and $\vec e_y$ are unit vectors in the $x$ and $y$ directions, respectively. After rephrasing the coordinate $x\vec e_x+y\vec e_y$ as $L(X^1 \vec e_x+  X^2\vec \tau)$ with $X^1,X^2\in[0,1]$, we can express the single-particle Hamiltonian as 
\begin{align}\label{H-tau}
H_0(\tau) &= \frac12 g^{ab}(\tau)D_a(\mathbf{A}) D_b(\mathbf{A}) 
\end{align}
with 
\begin{align}
g(\tau) &=\frac{1}{L^2\tau_2^2}
\left(
\begin{array}
{cc}
|\tau|^2	&	-\tau_1\\
-\tau_1		&	1
\end{array}
\right),
\end{align}
where the vector potential $\mathbf{A}=-\tau_2 L BX^2\vec e_x$, and the covariant derivative $D_a(\mathbf{A}) = -i\hbar \partial/\partial X^a+|e|A_a$. 
The inverse-mass-matrix $g(\tau)$ depends on the shape of the torus, 
which plays the role of a geometric metric~\cite{Read2009,Read2011,Scaffidi2017,Hoyos2012,Stern2019,Gromov2017a,Gromov2017b}. The total number of fluxes $N_{\phi}$ penetrating the torus is given by the Landau level degeneracy
$N_{\phi}=|\vec{L}_1 \times \vec{L}_2|/(2\pi \ell^2)$,
where the magnetic length $\ell=\sqrt{\hbar/(|e|B)}$ is taken as the length unit. The filling factor is defined as $\nu=N_p/N_\phi$.

We focus on the continuous geometric deformation generated by the Dehn twist operation on the torus, which corresponds to the adiabatic process $\vec \tau \rightarrow \vec \tau+\vec e_x$ as illustrated in Fig.~\ref{fig:torus}. Since the torus geometry after Dehn twist is equivalent to the original one, the physics of a topological order should be left unchanged \cite{Niu1985}.
As required by the principle of gauge invariance, we expect that the nearly degenerate ground-state manifold of a stable FQH phase
should evolve adiabatically in the whole process of Dehn twist, and each ground state should finally acquire a sector-dependent Berry phase. 
One advantage of our choice of Dehn twist is the potentially rich information contained in these Berry phases. First, the process of Dehn twist operation is equivalent to shearing the torus geometry, which is similar to applying a strain to a fluid. As a result, these Berry phases should reflect the viscosity response of FQH states~\cite{Avron1995,Read2009,Read2011}. Second, the Dehn twist operation coincides with $\mathcal{T}$-transformation in 1+1D conformal field theory \cite{CFT_book}, thus we expect to extract the topological properties of modular tensor category from these Berry phases~\cite{YZhang2012,WZhu2013,WZhu2014,Moradi2015}. Third, the adiabatic evolution of the FQH ground-state manifold itself can be used as a criterion for the robustness of FQH liquids, which so far has not been confirmed by proof-of-principle numerical evidence.

\subsection{Geometric phase}
In order to explicitly illustrate the Dehn-twist induced Berry phase, 
we first consider the $\nu=1/q$ Laughlin wave function in the topological sector $\alpha$ on the torus~\cite{Wen2012}:
\begin{align}\label{eq:Abelian-Wave}
\langle \{z_i\}| \Psi^\alpha;\tau\rangle &= \mathcal{N}(\tau) \prod_{i<j} \left[\frac{\theta_{11}(z_i-z_j|\tau)}{\eta(\tau)} \right]^q \times \nonumber\\
&f^\alpha_c\left(\left\lbrace Z \right\rbrace | \tau\right) e^{i\pi N_{\phi}\tau \sum_i [X_{2,i}]^2},
\end{align}
where $z_i=x_i+iy_i$ is the coordinate of the $i$th particle. $\alpha$ labels the $q$ degenerate Laughlin states, each of which corresponds to a fixed type of quasiparticle. The center-of-mass part of the wave function is described by $f_c(\left\lbrace Z \right\rbrace )$,
the relative part is captured by the Jacobi-theta function $\theta_{11}(z_i-z_j|\tau)$, and
the normalization prefactor $\mathcal{N}(\tau) = N_0 [ \sqrt{\tau_2} \eta^2(\tau)]^{N_{\phi}/2}$ with
$N_0$ a $\tau$-independent constant and $\eta$ the Dedekind's function. 
For the Laughlin wave function Eq.~(\ref{eq:Abelian-Wave}), we can analytically prove that, up to an $N_p$-dependent term which can be removed by a gauge transform, 
the Dehn-twist induced Berry phase is  
\begin{equation}\label{eq:berryphase}
  U_\alpha^{\mathcal{T}}= -\eta^H L^2 + 2\pi h_{\alpha} - 2\pi\frac{c}{24}, 
\end{equation}
where the Hall viscosity $\eta^H$, the topological spin $h_{\alpha}$, and the chiral central charge $c$ will be defined as below.  
Details of $f_c(\left\lbrace Z \right\rbrace )$, $\eta^H$, and the proof of Eq.~(\ref{eq:berryphase}) are given in the supplementary material A, 
in which we also derive the Dehn twist induced Berry phase for general multicomponent Halperin model wave functions.

\subsection{Hall viscosity response}
As we see in Eq.~(\ref{eq:berryphase}), Dehn twist induces a path-dependent term in the Berry phase, which depends on
the total Hall viscosity $\eta^H$ and the path length $L$. This term comes from the stress response,
which is generally nonzero for time-reversal symmetry breaking Hall liquids~\cite{Read2009,Scaffidi2017,Hoyos2012,Stern2019,Gromov2017a,Gromov2017b}.
Using Eq.~(\ref{eq:Abelian-Wave}),
we calculate the $\mathcal{T}$-path induced geometric Berry connection as (see supplemental materials Sec.~A.5)
\begin{align}\label{Berry-Connection}
\mathcal{A}_{\tau}&=i \langle \Psi;\bar{\tau}|\partial_{\tau}|\Psi;\tau\rangle = 
-\frac{q N_p}{8\tau_2},
\end{align}
which is $\alpha$-independent and gives an adiabatic phase
\begin{equation}\label{eq:viscosity}
\int_0^1\mathcal{A}_{\tau_1} \mathrm{d}\tau_1= \int_0^1(\mathcal{A}_{\tau} +\mathcal{A}_{\bar\tau})\mathrm{d}\tau_1=-\frac{q N_p}{4\tau_2}= -\hbar^{-1}\eta^H L^2.
\end{equation}
Several remarks should be clarified here: 
(i) $\eta^H$ describes how the internal spin degree of freedom \cite{Haldane2009,Haldane2011} of an FQH state responds to the geometric deformation. 
It is related to the ``topological shift'' $\mathcal{S}$ by $\eta^H= \frac{\hbar q N_p}{4\tau_2 |L|^2}= \frac{\hbar \nu}{8\pi \ell^2}\mathcal{S}$  \cite{Wen1992,Read2009}.
Usually $\mathcal{S}$ only appears on curved manifolds, such as the spherical geometry, which
modifies the particle-flux relation to $N_\phi=N_p \nu^{-1}-\mathcal{S}$. 
The appearance of $\eta^H$ in Eq.~(\ref{eq:berryphase}) again reflects that the geometric nature of Dehn twist operation.
(ii) The shift $\mathcal{S}$ is related to the so called ``orbital spin'' $\overline{s}$ of an FQH droplet via $\mathcal{S}=2\overline{s}$. The orbital spin includes contributions from both Landau orbital and guiding center orbital: $\overline{s}=\tilde{s}-\frac{s}{p}$, where $\tilde{s}$ is Landau orbital spin 
and $s/p$ is averaged guiding center spin \cite{YeJePark2014} 
(see supplementary material Sec.~D). We can define a guiding-center viscosity as $\eta^g=\frac{\hbar}{4\pi \ell^2}\frac{-s}{q}$.
(iii) The geometric phase is determined by the square of path length $L^2$, 
rather than the system area $|\vec L_1 \times \vec L_2|$, which reflects the path-dependent nature of the viscosity response.

\subsection{Modular response}
Besides the stress response, the Dehn twist, i.e., the $\mathcal{T}$-transformation, is expected to
encode topological information of the modular group on the torus~\cite{YZhang2012,Wen2012}.
Starting from Eq.~(\ref{eq:Abelian-Wave}), we can prove that (see details in supplemental material Sec.~A.4)
\begin{align}
\langle \{z_i\}| \Psi^\alpha;\tau+1\rangle&=&\langle \{z_i\}| \Psi^\alpha;\tau\rangle e^{2\pi i(h_\alpha-\frac{c}{24})} e^{i\frac{\pi q N_p^2}{12}}, 
\end{align}
which gives us the matrix representation of $\mathcal{T}$ under the basis of initial states as
\begin{align}
 \langle \Psi^\beta;\tau|\mathcal{T} |\Psi^\alpha;\tau\rangle =   \langle \Psi^\beta;\tau|\Psi^\alpha;\tau+1\rangle= T_{\alpha\beta} e^{i\gamma}.
 \label{eq8}
\end{align}
This relation indeed recovers the modular $T$-matrix
\begin{align}\label{eq:modular}
 T_{\alpha\beta} = \delta_{\alpha \beta} e^{i2 \pi (h_\alpha- \frac{c}{24})}.
\end{align}
Here, $h_\alpha$ is known as the ``topological spin'' of the topological sector $\alpha$,
and it qualifies the Berry phase obtained in the adiabatic self-rotation
of a quasiparticle.
$c$ is chiral central charge which determines the $1+1$D edge state structure 
within conformal field theory. 
Therefore, the modular matrix extracted from microscopic ground-state wave functions can characterize the underlying topological phase \cite{YZhang2012,WZhu2013,WZhu2014}.

In addition to the modular information, 
we also obtain a system-size dependent phase factor
\begin{align}\label{eq:gaugephase}
\gamma=\frac{\pi q N_p^2}{12}
\end{align}
in Eq.~(\ref{eq8}) (see supplemental material Sec.~A.4), which was overlooked in previous literatures~\cite{Zaletel2013}. 
However, it is necessary to fix this phase if we want to determine the central charge from microscopic wave functions~\cite{note2}.

\subsection{Gauge Transformation}
As numerical simulations of FQH problems are often implemented in the Landau level orbital basis, which will be clarified in the next section, here we derive the relation between the orbital bases before and after the Dehn twist shown in Fig.~\ref{fig:torus}(b). 
Before the Dehn twist, the two elementary magnetic translational operators 
$  \hat{t}_1 = \hat{t}(\frac{\vec{L}_1}{N_{\phi}}) $ and
$  \hat{t}_2 = \hat{t}(\frac{\vec{L}_2}{N_{\phi}}) $
on the rectangular torus satisfy $ \hat{t}_1 \hat{t}_2 = \hat{t}_2 \hat{t}_1 e^{i\frac{2\pi N_p}{N_{\phi}}}$ and act on the 
single-particle orbital basis as~\cite{Haldane1985,Bernevig2012}
\begin{equation}\label{eq:translation-basis}
\hat{t}_1 |m\rangle =  e^{i\frac{2\pi m}{N_{\phi}}} |m\rangle  ,\,\,\,\,
\hat{t}_2 |m\rangle = |m+1 \rangle,
\end{equation}
where $m=0,1,\cdots, N_\phi-1$ and $\hat{t}(\vec{r})$ is the general magnetic translational operator.
After the $\mathcal{T}-$transformation, the orbital basis should be the same as the initial one up to a gauge phase $\gamma_m$, i.e.,
$\mathcal{T} |m\rangle = |\overline{m}\rangle = e^{i\gamma_m} |m\rangle$,
where $|\overline{m}\rangle$ stands for the basis after the $\mathcal{T}-$transformation and satisfies
\begin{equation}\label{eq:translation-newbasis}
\hat{t}_1 |\overline{m}\rangle =  e^{i\frac{2\pi m}{N_{\phi}}} |\overline{m}\rangle ,\,\,\,\,
\hat{t} \left(\frac{\vec{L}_1+\vec{L}_2}{N_{\phi}}\right) |\overline{m}\rangle = |\overline{m+1} \rangle.
\end{equation}
A combination of Eqs.~(\ref{eq:translation-basis}) and (\ref{eq:translation-newbasis}) leads to $\gamma_{m+1}-\gamma_m=(2m+1)\frac{\pi}{N_{\phi}}$. 
Assuming $\gamma_m=Am^2+Bm+C$, we get
\begin{equation}\label{eq:gauge}
\gamma_m= \pi \frac{ m^2}{N_{\phi}} + C.
\end{equation}
In the many-body level, the total $U(1)$ gauge phase $\gamma'$ between two equivalent orbital-basis Fock states before and after the Dehn twist is simply the sum of Eq.~(\ref{eq:gauge}) over all occupied orbitals, i.e.,
\begin{align}\label{eq:gaugeoperator}
\hat{U}_g=\prod_{m\in{\rm occupied}}  e^{-i\gamma_m}|m\rangle \langle\bar{m}| 
\end{align}

An analytical derivation based on the real-space wave function of the orbital basis gives $C=0$ in Eq.~(\ref{eq:gauge}) (see supplemental material Sec.~A.2). However, in the following we set $C=\gamma/N_p$ instead, which, according to Eq.~(\ref{eq:gaugeoperator}), is equivalent to a gauge transform $|\Psi^\alpha;\tau+1\rangle\rightarrow e^{i\gamma}|\Psi^\alpha;\tau+1\rangle$. The advantage of this choice is that the system-size dependent $\gamma$ in Eq.~(\ref{eq8}) can be canceled by $C N_p$, such that Eq.~(\ref{eq:berryphase}) is obtained as a combination of Eqs.~(\ref{eq:viscosity}) and (\ref{eq8}). In this case, the Berry phase Eq.~(\ref{eq:berryphase}) is independent on $N_p$, containing only the pure geometric and topological terms.

\begin{figure*}[t]
	\includegraphics[width=0.79\textwidth]{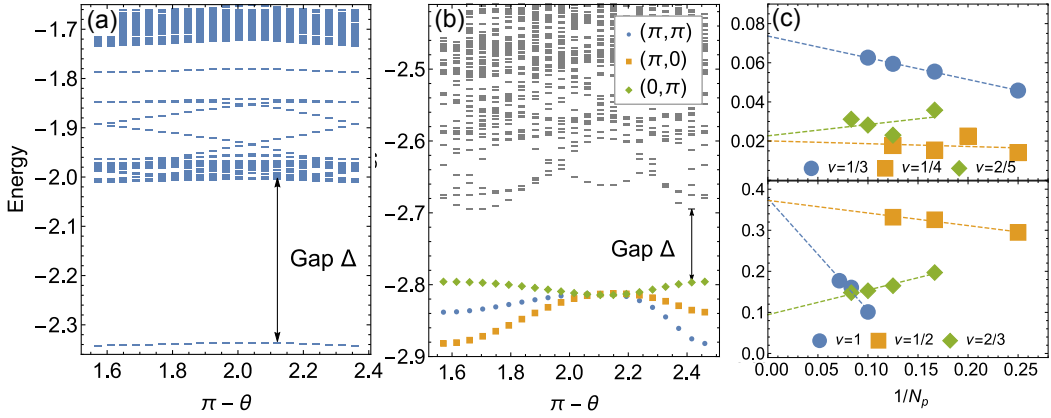}
	\caption{\textbf{Flow of low-energy spectra.} 
		The low-energy spectra of the Coulomb interaction as a function of the twist angle $\theta$ for (a) bosons at $\nu=1/2$ with $N_p=8$ and
		(b) bosons at $\nu=1$ with $N_p=12$. 
		The initial geometric conditions before the Dehn twist are chosen as $ |\vec L_1|=|\vec L_2|$ in (a) and $ |\vec L_1|=1.23|\vec L_2|$ in (b), with $\theta=\pi/2$. 
		The spectra are calculated in the irreducible Brillouin zone $(K_1,K_2)$ with $K_{1,2}=0,1,...,N_\phi/q-1$.
		(c) Finite-size scaling of the minimal energy gap $\Delta$ during the Dehn twist for the $\nu=1/2,1/3,1/4$ Laughlin states, $\nu=2/5,2/3$ hierarchy states and
		 $\nu=1$ Moore-Read state. The arrows in (a) and (b) indicate the minimal gap during the Dehn twist.
	} \label{fig:energy}
\end{figure*}

\section{Microscopic model}

We consider the Dehn twist driven evolution of a many-particle system under 
a translation-invariant two-body interaction
projected to the lowest Landau level. Such a Hamiltonian has the form of
\begin{equation}
H(\tau)=\frac{1}{2L_1 L_2\sin\theta}
\sum_\mathbf{q}V_\mathbf{q}:\hat{\rho}_{-\mathbf{q}}\hat{\rho}_\mathbf{q}: \label{eq:projected-Ham}
\end{equation}
where $V_\mathbf{q}$
is the Fourier transform of the interaction potential and $\hat{\rho}_\mathbf{q}$ is the projected density operator.
The standard second quantization procedure gives 
\begin{equation}
\hat{\rho}_\mathbf{q}=\int \mathrm{d}\mathbf{r}\,
e^{-i\mathbf{q}\cdot\mathbf{r}}\sum_{j_1,j_2}
\psi_{j_1}^*(\mathbf{r}) \psi_{j_2}(\mathbf{r}) a_{j_1}^\dagger a_{j_2},
\end{equation}
where the single-particle orbital can be taken as 
$\psi_{j}(x,y)=\left(\frac{1}{\pi^{1/2} L \ell}\right)^{1/2}  \sum_{k} e^{2\pi (j+k N_{\phi}) 
\frac{x+iy}{L}+i\frac{\pi \tau}{N_{\phi}}(j+kN_{\phi})^2} e^{-\frac{y^2}{2\ell^2}}$
with $ j =0,1,\cdots, N_{\phi}-1$, and $a^{\dagger}_j$ creates a particle in orbital $j$.
In the following, we choose the Coulomb interaction with $V_\mathbf{q}=\frac{2\pi}{|\mathbf{q}|}$ or Haldane's pseudopotentials.
We use exact diagonalization to calculate the energy spectrum and eigenstates of Eq.~(\ref{eq:projected-Ham})
at $\nu=p/q$ with coprime $p$ and $q$.
The full many-body symmetry can be factorized into a center-of-mass and a relative part \cite{Haldane1985,Bernevig2012},
thus each eigenstate is labeled by a two-dimensional momentum ${\bf K}=(K_1,K_2)$ with $K_{1,2}=0,1,\cdots,N_\phi/q-1$ in the irreducible Brillouin zone.

We use the discretization scheme to extract the Berry phase accumulated during Dehn twist~\cite{Read2011,Fremling2014}. 
To be specific, we parametrize the Dehn twist process in Fig.~\ref{fig:torus}(b) by the twist angle $\theta$ of the torus which varies continuously from $\pi/2$ to $\tan^{-1}(|\vec L_2(\theta=\pi/2)|/|\vec L_1|)$. The accumulated Berry phase $U_a^{\mathcal{T}} $ can then be defined as an integral of the
Berry connection $A(\theta)=i\langle\Psi_a(\theta)|\partial_\theta \Psi_a(\theta)\rangle$, i.e. $e^{iU_a^{\mathcal{T}}}= e^{i\int A(\theta) d\theta}$,
where $|\Psi_a(\theta)\rangle$ is the ground state of $H(\tau(\theta))$ [Eq.~(\ref{eq:projected-Ham})] in topological sector $a$ and $\tau=\tau(\theta)$ depends on $\theta$.
Dividing the path of $\theta$ into $M$ steps, we can get a discrete formula for $U_a^{\mathcal{T}} $ as
\begin{align} \label{eq:product}
e^{iU_a^{\mathcal{T}}} \simeq \langle \Psi_a(M)|\hat U_g |\Psi_a(M-1) \rangle 
\prod_{j=0}^{M-2} \langle\Psi_a(j+1)|\Psi_a(j)\rangle,
\end{align}
where $|\Psi_a(j)\rangle=|\Psi_a(\theta_j)\rangle$ with $\theta_j$'s evenly spaced along the path of $\theta$ ($j=0,1,\cdots,M\equiv 0$) and $\hat U_g$ is the gauge-relevant operator given by Eq.~(\ref{eq:gaugeoperator}) \cite{note4}. 
Note that global phases randomly returned by numerical diagonalization are automatically canceled in Eq.~(\ref{eq:product}). 
When using Eq.~(\ref{eq:product}), we inspect the wave-function overlap in each step of the Dehn twist 
to insure that the deformation is indeed performed adiabatically ($|\langle\Psi_a(j+1)|\Psi_a(j)\rangle|>0.99$).

\begin{table*}[t]
	\caption{In this table, we compare our numerical results with theoretical predictions. $s$ is the guiding center spin, being related to
		the guiding center Hall viscosity by $\eta^g=\frac{\hbar}{4\pi \ell^2}\frac{-s}{q}$.
		If the Landau-orbital part is included, $s$ directly determines the topological shift $\mathcal{S}$ of each FQH state,
		which corresponds to the total Hall viscosity $\eta^H=\frac{\hbar}{4}\frac{\nu}{2\pi \ell^2}\mathcal{S}$ \cite{Read2009}.
		The relationship between $s$ and $\mathcal{S}$ is given in supplemental materials Sec.~D.
		$h_a$ is the sector-dependent topological spin. See supplemental materials Sec.~C for detailed information about the topological sectors of each FQH state. Quantities with and without the superscript ``cal'' stand for numerically calculated results and theoretical values, respectively. We use parent Hamiltonians for the $(221)$ and $(332)$ Halperin states, otherwise we use the Coulomb interaction. 
	}
	\begin{ruledtabular}\label{tab:spin}
		\begin{tabular}{c|ccc|cc|c|cc}
			&   Laughlin & Laughlin &    Laughlin          &Hierarchy&     Hierarchy        & Moore-Read    & Halperin $(221)$ &  Halperin $(332)$   \\
			\hline
			$\nu=\frac{p}{q}$ & $\frac{1}{2}$&$\frac{1}{3}$&$\frac{1}{4}$& $\frac{2}{5}$&$\frac{2}{3}$&  $\frac22$&  $\frac23$  & $\frac25$   \\
			$\mathcal{S}$     &$2$           & $3$          &$4$          &$4$          & $3$        &  $2$  &    $2$      &    $3$    \\
			\hline
			$s$               & $-\frac{1}{2}$&$-1$         &$-\frac{3}{2}$&$-3$         & $-2$        &  $-1$  &  $-1$ & $-2$  \\
			$s^{\rm cal}$         & $-0.4997$    &$-0.9964$     &$-1.4469$     &$-2.9552$     & $-2.0840$  &  $-1.0320,-1.0246$& $-1.0499$  &  $-2.0033$ \\
			\hline
			$h_a-h_0$         & $\frac{1}{4}$&$\frac{1}{3}$ &$\frac{1}{8},\frac{1}{2}$&$-$ &$-$      &   $\frac{1}{2},\frac{3}{16}$&  $\frac{1}{3}$   &  $\frac{1}{5},\frac{2}{5}$ \\
			$h^{\rm cal}_a-h^{\rm cal}_0$     & $0.2500$     &$0.3333$      &$0.1250,0.5000$          &$0.2000,0.4000$&$0.3333$& $0.5000,0.1873$ &  $0.3333$  &  $0.2000,0.4000$
	\end{tabular}\end{ruledtabular}
\end{table*}

\section{Results}
\subsection{Flow of Energy Spectra}
As an isolated ground-state manifold in the whole process of the geometric deformation is a requisite for a well defined Berry phase, we first investigate the evolution of the low-energy spectra during the Dehn twist. Such an examination can reflect the stability of the FQH phase under the geometric deformation. Due to the relevance with realistic systems, we consider Coulomb interacting particles in what follows. 

Remarkably, we observe an impressive robustness of Abelian FQH states against the Dehn twist. A typical example of $\nu=1/2$ bosons is displayed in Fig.~\ref{fig:energy}(a). Here we choose a geometric path from a rectangular to its equivalent one (as shown in Fig.~\ref{fig:torus}). In this case, there is always a single ground state in the $(K_1,K_2)=(0,0)$ momentum sector, which we confirm has a large overlap with the Laughlin state and never mixes with other excited levels as the twist angle $\theta$ of the torus changes during the Dehn twist. For each system size, the energy gap separating the ground state and excited states is almost constant during the Dehn twist even for the generic Coulomb interaction [Fig. \ref{fig:energy}(a)]. A finite-size scaling of the minimal energy gap $\Delta$ in the process of Dehn twist suggests that the gap is very likely to survive in the thermodynamic limit  [Fig. \ref{fig:energy}(c)]. In addition, the minimum of the magnetoroton mode [at the bottom of excited levels in Fig.~\ref{fig:energy}(a)] only changes a little with $\theta$, indicating that not only the ground state but also the low-energy excitations are stable against the Dehn twist.

We observe similar robustness of the ground-state manifold for other bosonic and fermionic Coulomb ground states at $\nu=1/4, 1/3, 2/3$ and $2/5$, 
which correspond to the Abelian Laughlin, hierarchy, and Halperin states (see supplemental materials Sec.~C). In all of these cases, the single ground state in the irreducible Brillouin zone evolves adiabatically and never mixes with excited levels during the Dehn twist. The energy gap is also expected to be finite, as indicated by the finite-size scaling of the minimal gap during the Dehn twist [Fig. \ref{fig:energy}(c)].

For non-Abelian FQH states, there are multiple ground states in the irreducible Brillouin zone, which makes the spectral flow more complicated. To pursue small finite-size effects in the Coulomb ground states, we focus on bosons at $\nu=1$, where it has been confirmed that the Coulomb ground states are in the Moore-Read phase~\cite{zliuMR}. In this case, we again find remarkable robustness against the Dehn twist. Although the ground state in the $(K_1,K_2)$ sector evolves into the one in the $(K_1,K_1+K_2)$ sector after the Dehn twist, the three ground states are always approximately degenerate and well separated from other excited levels by a finite energy gap [Fig. \ref{fig:energy}(c)] in the whole spectra flow [Fig.~\ref{fig:energy}(b)].

\begin{figure}[h]
\includegraphics[width=0.48\linewidth]{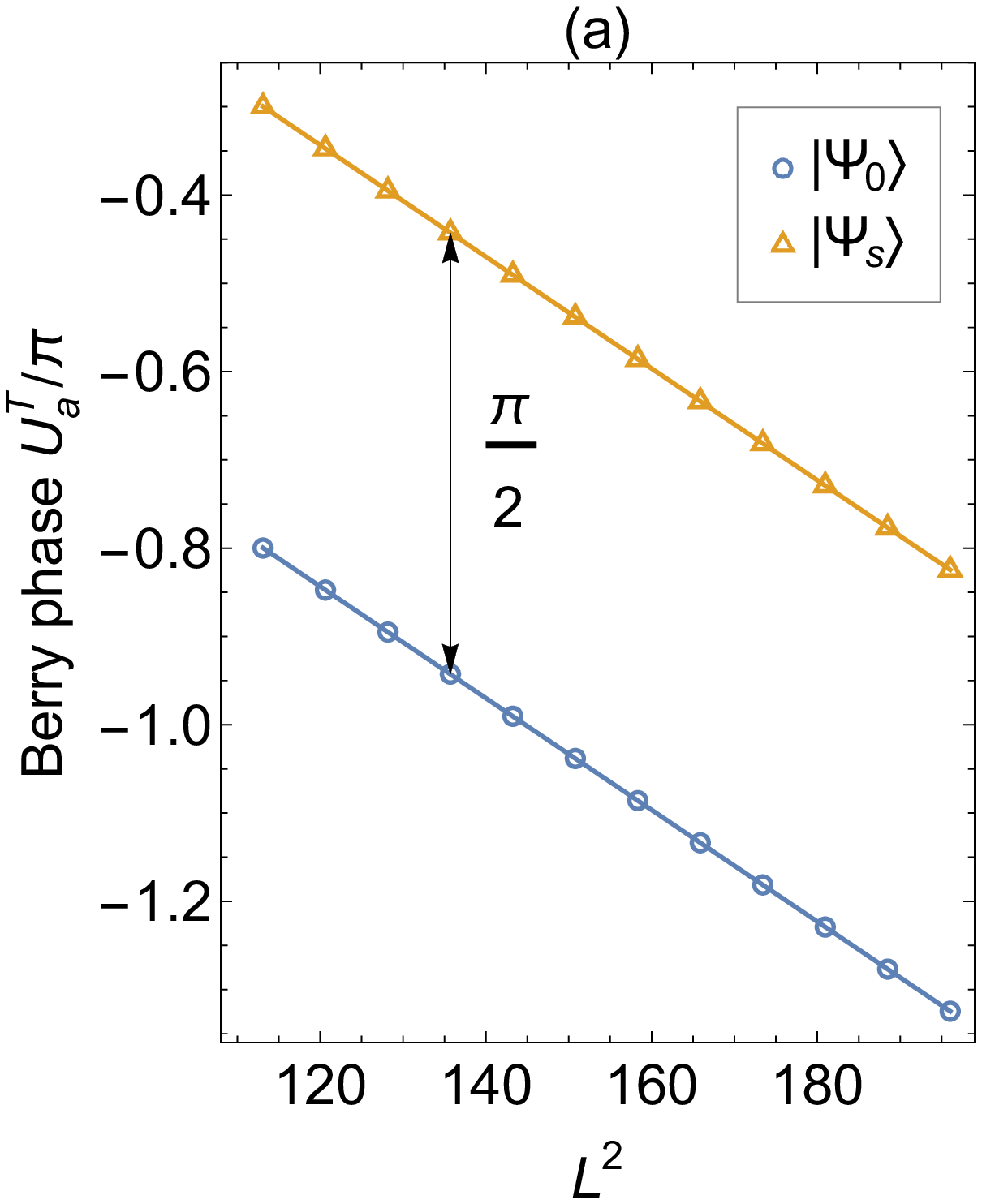}
\includegraphics[width=0.465\linewidth]{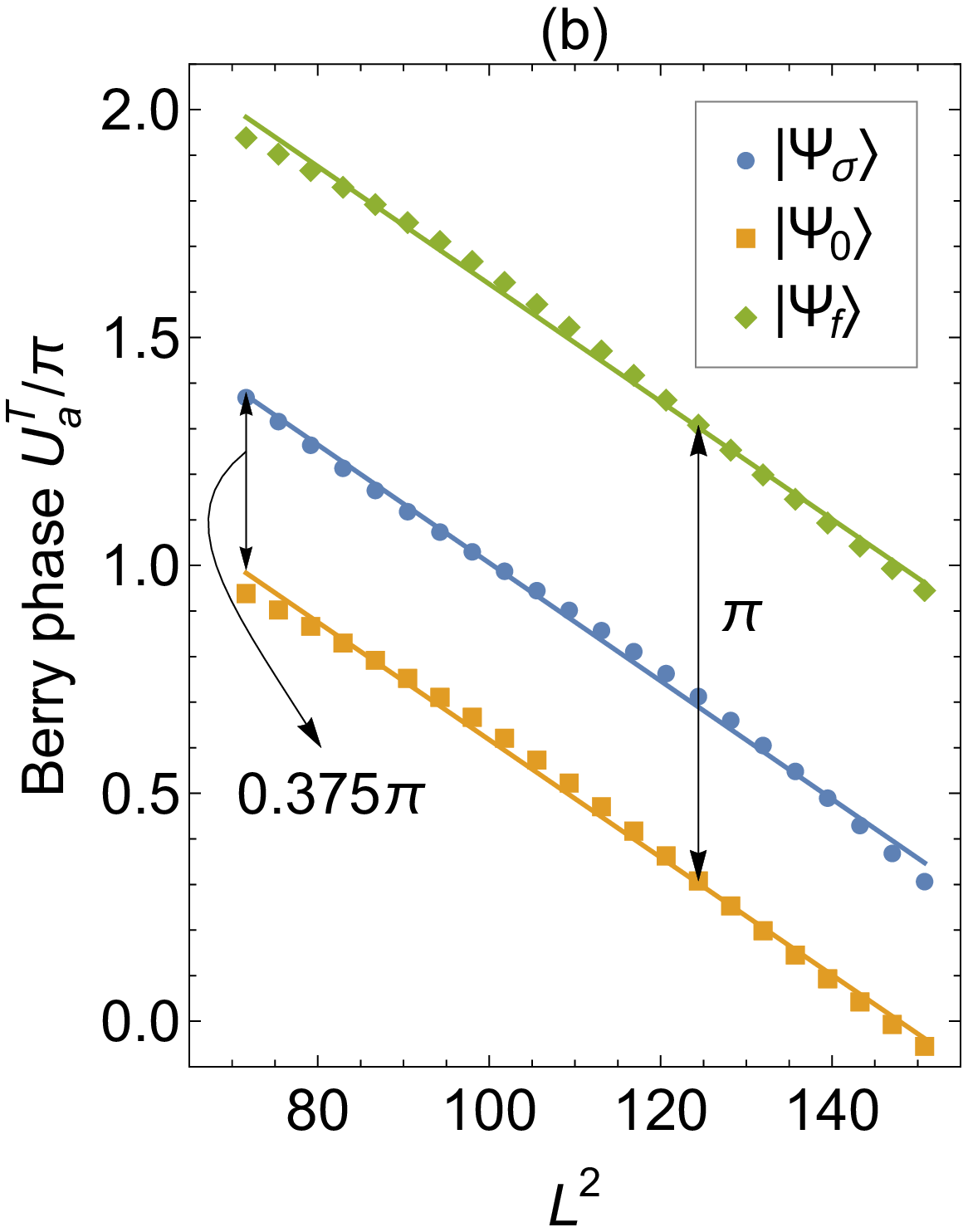}
\caption{\textbf{Accumulated Berry phase under Dehn-twist $\vec \tau\rightarrow \vec \tau+\vec{e}_x$}. 
Data are numerically calculated for Coulomb interacting bosons at (a) $\nu=1/2$ and (b) $\nu=1$, with $N_p=12$ in both cases.
Different topological sectors, with notations given in the main text, are distinguished by colors.
Fitting each curve into Eq.~(\ref{eq:berryphase}) allows us to extract the guiding-center spin and topological spin in different topological sectors.
} \label{fig:ut}
\end{figure}

\subsection{Berry phase and Hall viscosity}
Let us now turn to discuss the accumulated Berry phase under the Dehn-twist operation. 
For specific $N_p$, $\nu$, $N_\phi=N_p/\nu$ and topological sector $a$, we first numerically calculate the Dehn-twist induced Berry phase at a fixed length $L=|\vec L_1|$ of the torus.  We then vary $L$ around the square torus limit $L=\sqrt{2\pi N_\phi}$ to investigate the dependence of the Berry phase on $L$. We do these procedures in each topological sector $a$. Note that the torus area $|\vec L_1\times \vec L_2|=2\pi N_\phi$ is unchanged when we tune $L$.

Remarkably, for various Abelian and non-Abelian FQH states that we have studied, the numerically obtained Berry phase $U^{\mathcal{T}}_a$ in each topological sector $a$ behaves nicely as a linear function of $L^2$ in the window of $L$ stated above (Fig. \ref{fig:ut}), which is consistent with the prediction of Eq.~(\ref{eq:berryphase}). Thus we expect that the slope of the linear function $U^{\mathcal{T}}_a(L^2)$ is given by the sector-independent guiding center Hall viscosity $\eta^g=-\frac{\hbar }{4\pi \ell^2}\frac{s}{q}$ of the underlying FQH state, with $-\frac{s}{q}$ the averaged guiding center spin.
Physically, $s$ describes an emergent geometric response of a correlated composite boson (with $p$ particles in consecutive $q$ orbitals) and can be used as a topological quantum number to distinguish different FQH states~\cite{Haldane2011,YeJePark2014,Read2009}. The reason why we expect $\eta^g$ rather than the total $\eta^H$ in the slope extracted from numerical data is that we have projected the Hamiltonian into a single Landau level such that only the guiding-center part can be captured (see supplementary material Sec. D).
On the other hand, the sector-dependent topological spin $h_{a}$, describing
the phase obtained by quasiparticle $a$ spinned by $2\pi$, and the chiral central charge $c$ are expected to be encoded in the intercept of $U^{\mathcal{T}}_a(L^2)$ in the limit of $L\rightarrow 0$. In particular, the difference between the intercepts of $U^{\mathcal{T}}_a(L^2)$ and $U^{\mathcal{T}}_b(L^2)$ should give us the topological spin difference $h_{a}-h_{b}$. 

For Abelian states, the ground states in different topological sectors can be distinguished by their momenta $(K_1,K_2)$, thus we straightforwardly have $|\Psi_a\rangle=|\Psi(K_1,K_2)\rangle$, where $|\Psi(K_1,K_2)\rangle$ is the ground state from numerical exact diagonalization. Based on this, we calculate the Berry phase $U^{\mathcal{T}}_a$, and indeed extract guiding center spin and sector-dependent topological spin that are very close to their theoretical values in pertinent FQH phases. For instance, we get $s\approx -0.4997$ for the two degenerate Coulomb ground states of bosons at $\nu=1/2$ and the intercept difference gives $\Delta h\approx 0.2500$ [Fig. \ref{fig:ut}(a)]. This matches the $\nu=1/2$ bosonic Laughlin state which carries $s=-1/2$ and has two types of quasiparticles with $h_0=0$ ($a=0$ vacuum) and $h_{\rm s}=1/4$ ($a={\rm s}$ semion), respectively~\cite{YeJePark2014,Cincio2013,WZhu2013}. 
We have also explored other Abelian FQH states corresponding to the Laughlin states at $\nu=1/3,1/4$, hierarchy states at $\nu=2/5,2/3$, and Halperin states at $\nu=2/3,2/5$ (see supplemental materials Sec.~B). We summarize these results in Tab.~\ref{tab:spin}, where all of the numerically extracted guiding center spin and topological spin are consistent with theoretical predictions based on Jack polynomials or model wavefunction calculations \cite{YeJePark2014}.

For non-Abelian states, we need to appropriately superpose the ground states $|\Psi(K_1,K_2)\rangle$ obtained from numerical exact diagonalization to construct the state $|\Psi_a\rangle$ in a specific topological sector $a$. Here we take the $\nu=1$ Coulomb interacting bosons in the Moore-Read phase as an example. In this case, the three numerical ground states are in the $(K_1,K_2)=(\pi,0),(0,\pi)$ and $(\pi,\pi)$ momentum sectors. The Moore-Read phase has three types of quasiparticles: the vacuum $a=0$, the fermionic anyon $a=f$ and the Ising anyon $a =\sigma$. In particular, the Ising anyon $\sigma$ carries non-Abelian braiding statistics which can lead to potential applications in topological quantum computation \cite{Kitaev2003,Nayak}. Based on the symmetry analysis, $|\Psi_a\rangle$ and $|\Psi(K_1,K_2)\rangle$ are related via
$|\Psi_{\sigma}\rangle=|\Psi(0,\pi)\rangle$ and $|\Psi_{0,f}\rangle=\frac{1}{\sqrt{2}}(|\Psi(\pi,\pi)\rangle \pm e^{i\varphi}|\Psi(\pi,0)\rangle)$, where $\varphi$ 
is chosen to guarantee that $|\Psi_{0}\rangle$ and $|\Psi_{f}\rangle$ are minimally entangled states~\cite{YZhang2012,WZhu2013,WZhu2014} with respect to the bipartition of all $N_\phi$ Landau level orbitals (see supplementary material Sec.~E). 
Similar to Abelian cases, we find that the Dehn-twist induced Berry phase of each such constructed $|\Psi_a\rangle$ also matches a linear dependence on $L^2$ for $L$ around the square torus limit [Fig. \ref{fig:ut}(b)].
The extracted guiding center spin is $s\approx-1.0320$ and $-1.0246$ for $|\Psi_{\sigma}\rangle$ and $|\Psi_{0,f}\rangle$, respectively, which is almost sector-independent and very close to the theoretical value $s=-1$ in the Moore-Read phase.
The topological spin of $f$ and $\sigma$ are respectively determined by $h_f-h_0\approx 0.5000 $ and $h_{\sigma}-h_0\approx0.1873$,
being consistent with expected ``fermionic'' and ``Ising'' statistics of quasiparticle $f$ and $\sigma$~\cite{Kitaev2006,WZhu2015}.

\subsection{Chiral central charge and edge physics}
In the vacuum sector $a=0$, the topological spin $h_a=0$ such that the intercept of $U^{\mathcal{T}}_{a=0}(L^2)$ is solely contributed by the chiral central charge $c$. In this case, we can investigate the edge structure of an FQH state which is determined by $c$. 
As notable examples, we first consider the fermionic model ground state at $\nu=1/3$, and its particle-hole conjugate at $\nu=2/3$. 
Working in the vacuum sector, we extract the central charge of the $\nu=1/3$ ground state as $c\approx 1.01595$ (Fig.~\ref{fig:centralcharge}), which is close to the theoretical value $c=1$ for the $\nu=1/3$ Laughlin state (see supplementary material Sec.~B for details). Physically, $c=1$ means that the edge state is a single chiral bosonic field, being consistent with the well known edge structure of the $\nu=1/3$ Laughlin state.  By contrast, there are multiple scenarios of the edge physics of the particle-hole conjugate at $\nu=2/3$. One possibility is that the edge current is carried by two chiral $\nu=1/3$ edge states~\cite{Chang1992,Beenakker1990}. 
However, it has been debated that the $\nu=2/3$ state should harbor
two counter-propagating $\nu = 1$ and $\nu = 1/3$ edge modes and edge reconstruction could occur in this hole-conjugate FQH state \cite{Kane1994}.  
The difference between the two scenarios above is that 
the former hosts $c=2$, while the latter has $c=0$ due to the counter-propagating nature.
As shown in Fig. \ref{fig:centralcharge}, we obtain $c\approx 0.0159$ for the $\nu=2/3$ Coulomb ground state within very high accuracy.
This result unambiguously points to the counter-propagating picture \cite{Kane1994}
and is also consistent with the recent shot noise measurements and other experiments
\cite{Bid2010,Heiblum2014,Heiblum2019}.
In addition, we identify $c=1$ for the model Halperin $(333)$ state at $\nu=1/3$,
which suggests its effective edge theory to be equivalent to the Laughlin $\nu=1/3$ state.
In this sense, our approach offers a guide to explore the edge physics of existing FQH effects.

\begin{figure}[htb]
	\includegraphics[width=0.48\linewidth]{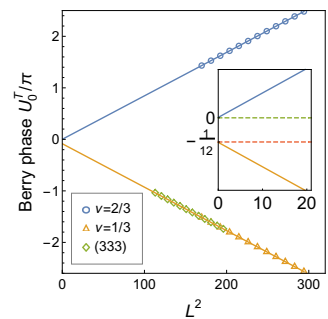}
	\caption{\textbf{Chiral central charge.}
		Linear extrapolation of the Berry phase $U^{\mathcal{T}}_{a=0}(L^2)$ towards $L=0$ 
		for the model ground state at $\nu=1/3$, its particle-hole conjugate at $\nu=2/3$.
		The intercept of the Berry phase in the vacuum sector $a=0$ gives 
		the chiral central charge $c\approx 1.01595$ for the $\nu=1/3$ ground state
		and $c\approx 0.0159$ for its particle-hole conjugate at $\nu=2/3$. 
		Similarly, we identify $c\approx 0.9818$ for the model Halperin $(333)$ state.
	} \label{fig:centralcharge}
\end{figure}

\begin{figure}[htb]
	\includegraphics[width=0.48\textwidth]{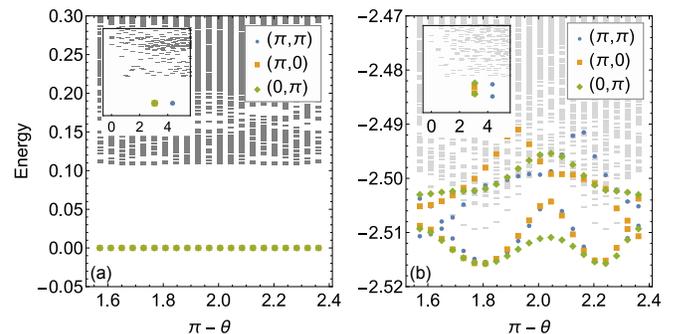}
	\caption{\textbf{Energy spectra at $\nu=5/2$.}
		The low-energy spectra of (a) the three-body parent Hamiltonian of the Pf state and 
		(b) the two-body Coulomb interaction at $\nu=1/2$ in the second Landau level as a function of the twist angle $\theta$ for $N_p=12$ electrons. 
		The lowest energy levels living in momentum sector $(0,\pi),(\pi,0),(\pi,\pi)$ are labeled by colors. The insets show the energy spectra at $\theta=\pi/2$ as a function of $|{\bf K}|$.
	} \label{fig:Pfenergy}
\end{figure}

\section{Discussion}
Apart from extracting topological and geometric quantum numbers of the underlying FQH state, our Dehn-twist approach also provides a distinctive viewpoint to inspect the stability of an FQH phase. In some cases such stability cannot be guaranteed by only studying finite-size samples with a fixed torus shape. 
Here we use the energy spectral flow under the Dehn twist as a criterion. As required by the gauge transformation, such an energy spectral flow 
is expected to maintain the ground-state degeneracy without level crossings with excited levels if the underlying FQH phase is really robust~\cite{Niu1985}.
The results shown in Figs.~\ref{fig:energy}(a) and \ref{fig:energy}(b) satisfy this requirement.
However, we also notice a striking counterexample for Coulomb interacting electrons at $\nu=5/2$ ($\nu=1/2$ in the second Landau level). In this case, 
while the nature of the ground state is still under debate, there is one possibility that the ground-state manifold consists of the non-Abelian Pfaffian (Pf) and anti-Pfaffian (aPf) states that are degenerate in the thermodynamic limit~\cite{Moore1991,Greiter1991,Morf1998,Haldane2000,Levin2007,Morf2010,HaoWang2009,Peterson2008}. As both the Pf and aPf states are three-fold degenerate in the irreducible Brillouin zone on the torus, the total ground-state degeneracy in the irreducible Brillouin zone is expected to be six-fold in this case.
Some numerical attempts indeed reported the observation of six low-lying states at $\nu=5/2$ on the torus of special shape~\cite{HaoWang2009,Peterson2008}.
However, we find that this feature is not stable under the Dehn twist.
As shown in Fig. \ref{fig:Pfenergy}(b), while there are six low-lying states on the rectangular torus \cite{HaoWang2009}, three of them evolve into the higher-energy spectrum during the Dehn twist, making the Pf-aPf interpretation questionable. 
It is in sharp contrast to the case of the particle-hole symmetry breaking three-body parent Hamiltonian of the (anti-)Pfaffian state, 
for which the (anti-)Pfaffian state is always the zero-energy ground state regardless of the torus shape [Fig. \ref{fig:Pfenergy}(a)].
Moreover, Ref.~\cite{Peterson2008} claimed that a quantum-well model with a finite layer-width could stabilize Pf and aPf states.
Unfortunately, we observe similar level crossing with excited levels in the spectral flow of that model also.
Thus, our calculations suggest that, compelling evidence on the torus geometry for Pf and aPf states at $\nu=5/2$ is still far from conclusive (see discussion in supplemental materials Sec.~B.5).

\section{Conclusion and Outlook}
In this work, we have presented a systematic scheme based on the Dehn-twist deformation on the torus geometry 
to identify the topological orders of fractional quantum Hall (FQH) liquids.
With a gauge fixing procedure, we analytically derive the correct formula of the geometric Berry phase accumulated during the Dehn twist. 
This formula explicitly relates the geometric response of FQH liquids to their universal properties, 
such as the Hall viscosity and the topological spin. We then verify this formula in various microscopic models of Abelian and non-Abelian FQH liquids beyond model wavefunctions, demonstrating the potential of our scheme as a diagnosis of the topological order in a generic FQH state without prior knowledge.
Motivated by the requirement of a well defined geometric Berry phase, we also suggest a separated ground-state manifold from excited levels in the whole process of geometric deformation as an indispensable criterion to justify the stability of an FQH phase.

Our approach opens up several future directions deserving further exploration.
We mostly focus on FQH states in the lowest Landau level in this work. Considering that a series of FQH effects are also observed in higher Landau levels, we believe that our Dehn-twist protocol can shed light on the stability of those FQH states and their difference from the lowest-Landau-level FQH states from the aspect of geometric response. Moreover, in order to deepen our understanding of the interplay between topology and geometry, it is instructive to investigate how the geometric response of FQH liquids is affected by the breaking of the rotational invariance, such as in the cases of anisotropic FQH states~\cite{Haldane2011,Papic2012} and FQH nematic phases~\cite{YLiu2013,Samkharadze2015}.
Furthermore, it would be interesting to adjust our Dehn-twist protocol to lattice systems, such that it can be applied to the broad class of lattice topological states such as fractional Chern insulators~\cite{parameswaran2013fractional,liu2013review,Titusreview}.


\textbf{Acknowledgements.---} 
L.D.H. and W.Z. thanks Tiansheng Zeng for simulating discussion,
and Bo Yang, Chong Wang, and Jie Wang for helpful discussions.
L.D.H. and W.Z. are supported by project 11974288 from NSFC and the foundation of Westlake University.
Z.L. is supported by project 11974014 from NSFC.
F.D.M.H. is supported by DOE grant No. DE-SC0002140. 
D.N.S is supported by National Science Foundation Grants PREM
DMR-1828019 and by the Princeton MRSEC through
the National Science Foundation Grant DMR-1420541.




\bibliographystyle{apsrev4-1}


%


\clearpage
\widetext
\appendix
\begin{appendices}

In this supplemental material, we provide more details of the calculation and results to support the discussion in the main text.
In Sec. A, we show a detailed derivation of geometric Berry phase shown in the main text. This section includes five subsections.
In Sec. B, we show the geometric deformation is independent on the selected twisted path. 
In Sec. C,  we present more numerical results at various filling factors which were not shown  in the main text. This section includes six subsections. 
In Sec. D, we address the relation between the current work and the previous studies. 
In Sec. E, we discuss how to make the gauge choice in the bosonic Moore-Read state.

\section{Theoretical Derivation for Abelian FQH states}
\subsection{Landau Level on Torus and Theta Function}
Considering an electron on a torus geometry with a uniform magnetic field perpendicular to 
the surface:
\begin{align}\label{H-tau-s}
H_0(\mathbf{A},\tau) &= \frac12 g^{ab}(\tau)D_a(\mathbf{A}) D_b(\mathbf{A}) , \\
g(\tau) &=\frac{1}{L^2\tau_2^2}
\left(
\begin{array}
{cc}
|\tau|^2	&	-\tau_1\\
-\tau_1		&	1
\end{array}
\right).
\end{align}
Where $D_a(\mathbf{A})=-i\hbar\partial/\partial X^a+|e|A_a$ and $\mathbf{A}=(-\tau_2 L^2 BX^2,0)$ are the covariant derivative and vector potential respectively. 
The ground states of Eq. \ref{H-tau-s} are $N_\phi$-fold degenerated:
\begin{equation}\label{Landau-level}
\Psi_m(X^1,X^2|\tau) = \frac{1}{\sqrt{\pi^{1/2}L\ell}}\mathrm{e}^{i\pi N_{\phi}\tau [X^2]^2}
\theta_{\frac{m}{N_\phi}}(N_\phi z/L|N_\phi \tau)
\end{equation}
where $N_\phi=\frac{\vec{L}_1\times\vec{L}_2}{2\pi\ell^2}=\frac{\tau_2 L^2}{2\pi\ell^2}$ is the total flux through the torus, $\ell=\sqrt{\hbar/|e|B}$ is the magnetic length,
and $z=x+iy=L(X^1+\tau X^2)$ is the complex coordinate of electron.
$\theta_{m}(z|\tau)$ is the theta function:
\begin{equation}
\theta_{\alpha}(z|\tau)=\sum_{n\in Z} \exp{\left( i\pi \tau (n+\alpha)^2 +i2\pi(n+\alpha)z \right) }.
\end{equation}
\subsection{Dehn twist and Gauge Transformation} \label{G-T}
Now, let us consider $\mathcal{T}$ transformation $\tau\rightarrow\tau+1$ and notice it implies an underlying coordinate transformation
$z=L(X^1+\tau X^2)=L(X'^1+(\tau+1) X'^2)$:
\begin{align}\label{H-tau1-s}
H_0(\mathbf{A}',\tau+1) &= \frac12 g^{ab}(\tau+1)D'_a(\mathbf{A}') D'_b(\mathbf{A}') , \\
\nonumber
g(\tau+1) &=\frac{1}{L^2\tau_2^2}
\left(
\begin{array}
{cc}
|\tau+1|^2	&	-\tau_1-1\\
-\tau_1-1		&	1
\end{array}
\right)
\end{align}
and the least Landau level wave function:
\begin{equation}\label{Landau-level-1}
\Psi_m(X'^1,X'^2|\tau+1) = \frac{1}{\sqrt{\pi^{1/2}L\ell}}\mathrm{e}^{i\pi N_{\phi}(\tau+1) [X'^2]^2}
\theta_{\frac{m}{N_\phi}}(N_\phi z/L|N_\phi (\tau+1)).
\end{equation}
Where $D'_a(\mathbf{A}')=-i\hbar\partial/\partial X'^a+|e|A'_a$ and $\mathbf{A}'=(-\tau_2 L^2 BX'^2,0)$ are the covariant derivative and vector potential in $(X'^1,X'^2)$. 
Note that we can't compare Eq. \ref{Landau-level-1} with Eq. \ref{Landau-level} since they are in different coordinate frame.
Therefore $H_0(\mathbf{A}',\tau+1)$ should be rewritten in $(X^1,X^2)$. Note that $g^{ab}(\tau+1)$ can be decomposed:
\begin{equation}\label{g-tau+1}
\left.
g(\tau+1)=\frac{1}{L^2\tau_2^2}
\left(
\begin{array}
{cc}
1	&	-1\\
0		&	1
\end{array}
\right)
g(\tau)
\left(
\begin{array}
{cc}
1	&	0\\
-1		&	1
\end{array}
\right)
\right.
\end{equation}
and 
\begin{equation}\label{d-transformation}
\left.
\left(
\begin{array}
{c}
D_1(\tilde{\mathbf{A}})	\\
D_2(\tilde{\mathbf{A}})	
\end{array}
\right)
=
\left(
\begin{array}
{cc}
1	&	0\\
-1		&	1
\end{array}
\right)
\left(
\begin{array}
{c}
D'_1(\mathbf{A}')	\\
D'_2(\mathbf{A}')	
\end{array}
\right)
\right.
\end{equation}
With $\tilde{\mathbf{A}}=(-\tau_2 L^2 B X'^2,\tau_2 L^2 B X'^2)=(-\tau_2 L^2 B X^2,\tau_2 L^2 B X^2)$.
By substituting Eq. \ref{g-tau+1} and \ref{d-transformation} into \ref{H-tau1-s}, we find that $H_0(\mathbf{A}',\tau+1)$ can be rewritten in coordinate $(X^1,X^2)$:
\begin{equation}\label{H-tau1-s-re}
H_0(\mathbf{A}',\tau+1) =H_0(\tilde{\mathbf{A}},\tau)= \frac12 g^{ab}(\tau)D_a(\tilde{\mathbf{A}}) D_b(\tilde{\mathbf{A}})  
\end{equation}
with $D_a(\tilde{\mathbf{A}})=-i\hbar\partial/\partial X^a+|e|\tilde{A}_a$. Now one can find a gauge transformation $\hat{\mathcal{U}}=\mathrm{e}^{-i\pi N_{\phi}[X^2]^2}$\cite{Wen2012}:
\begin{equation}\label{D-gauge}
D_a(\mathbf{A})=\hat{\mathcal{U}}^\dagger D_a(\tilde{\mathbf{A}})\hat{\mathcal{U}}
\end{equation}
thus Eq. \ref{H-tau-s} and Eq. \ref{H-tau1-s-re} are equivalent except a gauge transformation:
\begin{equation}\label{H-gauge}
H_0(\mathbf{A},\tau)=\hat{\mathcal{U}}^\dagger H_0(\tilde{\mathbf{A}},\tau)\hat{\mathcal{U}}.
\end{equation}
Using Eq. \ref{H-tau-s} \ref{Landau-level} \ref{Landau-level-1} \ref{H-tau1-s-re} and \ref{H-gauge} we have:
\begin{equation}\label{Landau-level-gauge}
\hat{\mathcal{U}}\Psi_m(X'^1,X'^2|\tau+1) = \mathrm{e}^{i\pi\frac{m^2}{N_\phi}}\Psi_m(X^1,X^2|\tau)
\end{equation}
Note that it is consistent with the result Eq. 13 in the main text. In Eq. \ref{Landau-level-gauge}, we have used this relation:
\begin{eqnarray}
\nonumber
\theta_{\frac{m}{N_\phi}}(N_{\phi}z|N_{\phi}(\tau+1)) &=& \sum_n \mathrm{e}^{i\pi N_{\phi} \tau (n+\frac{m}{N_{\phi}})^2 + i2 \pi (n+\frac{m}{N_{\phi}})N_{\phi}z 
	+i\pi N_{\phi} (n+\frac{m}{N_{\phi}})^2} \\
\nonumber
&=& e^{i\pi \frac{m^2}{N_{\phi}}}\sum_n (-1)^{nN_\phi}\mathrm{e}^{i\pi N_{\phi} \tau (n+\frac{m}{N_{\phi}})^2 + i2 \pi (n+\frac{m}{N_{\phi}})N_{\phi}z} \\
&=& e^{i\pi \frac{m^2}{N_{\phi}}}\theta_{\frac{m}{N_\phi}}(N_{\phi}z|N_{\phi}\tau) ~~~~~~~~~(N_\phi \text{ is even}).
\end{eqnarray}

\subsection{ FQH Wave Functions}
Now we consider the many-body problem. Starting with the multilayer FQH state (we only show the single-component case in the main text), we can derive the wave function on torus in terms of the theta function\cite{Fremling2014,Wen2012}:
\begin{eqnarray}\label{Abelian-Wave}
\nonumber
\Psi^{\bm{\alpha}}\left(\left\lbrace z_i^I \right\rbrace| \tau\right) &=& \mathcal{N}(\tau)
f_c(\left\lbrace Z^I \right\rbrace ) f_r( \left\lbrace z_i^I \right\rbrace ) \exp{\left\lbrace i\pi \tau N_{\phi} \sum_{I,i}\left( \frac{y_i^I}{L\tau_2}\right)^2\right\rbrace} \\
\nonumber
f_r\left( \left\lbrace z_i^I \right\rbrace| \tau \right) &=& \left\lbrace \prod_{I<J} \prod_{i,j} \eta^{-K_{IJ}}(\tau) \theta^{K_{IJ}}_{11}(z_i^I/L-z_j^J/L|\tau) \right\rbrace
\left\lbrace \prod_{I} \prod_{i<j} \eta^{-K_{II}}(\tau) \theta^{K_{II}}_{11}(z_i^I/L-z_j^I/L|\tau) \right\rbrace \\
f_c\left(\left\lbrace Z^I \right\rbrace | \tau\right) &=& \eta^{-\kappa}(\tau) f^{(\bm{\alpha},\bm{\eta})}(\bm{Z}/L|\tau)
\end{eqnarray}
where 
\begin{equation}
\theta_{11}(z|\tau)=\sum_{n\in \mathbb{Z}} \exp{\left\lbrace i\pi\tau (n+\frac12)^2 +i2\pi (n+\frac12)(z+\frac12) \right\rbrace }
\end{equation}
is the odd Jacobi-theta function which satisfy $\theta_{11}(-z|\tau)=-\theta_{11}(z|\tau)$ corresponds to the factor $(z_i-z_j)^m$ in Laughlin wave function, 
$f_r\left( \left\lbrace z_i^I \right\rbrace| \tau \right)$ is the relative part and the multi-dimension theta function $f^{(\bm{\alpha},\bm{\eta})}(\bm{Z}|\tau)$ in the CM function
$f_c\left(\left\lbrace Z^I \right\rbrace | \tau\right)$ is 
\begin{equation}\label{multi-theta}
f^{(\bm{\alpha},\bm{\eta})}(\bm{Z}|\tau) = \sum_{\bm{n}\in\mathbb{Z}^{\kappa}} 
\exp{ \left\lbrace i\pi(\bm{n}+\bm{\alpha}+\bm{\eta})^TK\tau(\bm{n}+\bm{\alpha}+\bm{\eta}) +i2\pi(\bm{n}+\bm{\alpha}+\bm{\eta})^TK(\bm{Z}-\bm{\eta}) \right\rbrace  }.
\end{equation}
The $\tau$-dependent Dedekind's $\eta$-function in Eq. \ref{Abelian-Wave} is given by:
\begin{equation}
\eta(\tau)=q^{1/24}\prod_{n=1}^{\infty}(1-q^n)|_{q=e^{i2\pi\tau}}
\end{equation}
and the normalization factor:
\begin{equation}
\mathcal{N}(\tau) = N_0 \left[ \sqrt{\tau_2} \eta(\tau)^2 \right]^{\frac{1}{2} \bm{\kappa}^T \bm{N}} 
\end{equation}
with $N_0$ an area-dependent constant.
For any given $K$ matrix with $dim(K)=\kappa$, $K_{IJ}$ is the enrties, $\bm{\kappa}=(K_{11},K_{22},\cdots,K_{\kappa\kappa})^T$ is the diagonal elements of $K$.
$\bm{N}=\left( N^1,N^2,\cdots,N^{\kappa}\right)^T $ where $N^I$ is the number of electrons in the $I$th layer, and $\bm{Z}=(Z^1,Z^2,\cdots,Z^\kappa)^T$ with
$Z^I=\sum_i z^I_i$ is the central mass coordinates of the $I$th layer and $z^I_i=L(X^{I1}_i+\tau X^{I2}_i)$ is the $i$th electron in the $I$th layer. The vectors used in \ref{multi-theta} 
are given by:
\begin{eqnarray}\label{alpha-eta}
\nonumber
\bm{\eta}    &=& K^{-1}\bm{\kappa}/2 ~~\text{(for fermion)}\\
\bm{\eta}    &=& \bm{0} ~~~~~~~~~~~~~~~\text{(for boson)}.
\end{eqnarray}
and $K\bm{\alpha}$ is the coset lattice $Z^\kappa/KZ^\kappa$ which only have $|\det(K)|$ independent vectors indicate $|\det(K)|$-fold degenerate on torus\cite{Wen1993,Fremling2014,Niu1990}.
For Laughlin $\nu=1/q$ state, its $K$ matrix is $K=q$ with coset lattice $Z/KZ=\left\{K\bm{\alpha}|0,1,\cdots,q-1\right\}$ correspond to the $q$-fold degeneracy. And for Halperin $(mmn)$ state, its $K$ matrix is:
$$
K=
\left(
\begin{array}{cc}
m & n \\
n & m \\
\end{array}
\right),
$$
and its coset lattice are enclosed by the parallelogram spanned by two vectors $(m,n)$ and $(n,m)$, the number of independent vectors is equal to the area of the parallelogram, i.e. $|\det{(K)}|=m^2-n^2$. If we consider Halperin $(332)$ state, the coset lattice is $Z^2/KZ^2=\left\{K\bm{\alpha}|(0,0),(1,1),(2,2),(3,3),(4,4)\right\}$. Once we got the coset lattice vector $K\bm{\alpha}$, the vector $\bm{\alpha}$ in wavefunction can be obtained by acting $K^{-1}$ on the left-hand side of $K\bm{\alpha}$, i.e. $\left\{\bm{\alpha}|0,1/q,\cdots,(q-1)/q\right\}$ for Laughlin $\nu=1/q$ state and $\left\{\bm{\alpha}|(0,0),(1/5,1/5),(2/5,2/5),(3/5,3/5),(4/5,4/5)\right\}$ for Halperin $(332)$ state.

\subsection{Dehn twist and Modular information}
Now, considering a special modular transformation $\left\lbrace \mathcal{T}:\tau \rightarrow \tau + 1 \right\rbrace $, 
similar to the single electron case, we should introduce a many-electron gauge transformation:
\begin{equation}\label{many-body-gauge}
\hat{\mathcal{U}}_g = \exp{\left\lbrace i\pi N_{\phi} \sum_{I,i}\left( \frac{y_i^I}{L\tau_2}\right)^2\right\rbrace}.
\end{equation}
Using Eq. \ref{Abelian-Wave}-\ref{many-body-gauge}, we have:
\begin{eqnarray}\label{tau+1-wave}
\nonumber
\hat{\mathcal{U}}_g\Psi^{\bm{\alpha}}\left(\left\lbrace z_i^I \right\rbrace| \tau+1\right)
&=&  \hat{\mathcal{U}}_g\mathcal{N}(\tau+1) f_c(\left\lbrace Z^I \right\rbrace|\tau+1 ) f_r( \left\lbrace z_i^I \right\rbrace|\tau+1 ) 
\mathrm{e}^{-i\pi N_{\phi} (\tau+1) \sum_{I,i}\left( y_i^I/L\tau_2\right)^2} \\
\nonumber
&=&  \mathcal{N}(\tau) f_c(\left\lbrace Z^I \right\rbrace|\tau ) f_r( \left\lbrace z_i^I \right\rbrace|\tau ) 
\mathrm{e}^{i\pi N_{\phi} \tau \sum_{I,i}\left( y_i^I/L\tau_2\right)^2}
\mathrm{e}^{\frac{1}{12}i\pi (\bm{N}^TK\bm{N}-\kappa )} \mathrm{e}^{i2\pi h_{\bm{\bm{\alpha}}}} \\
&=& \Psi^{\bm{\alpha}}\left(\left\lbrace X^{1}_{I,i},X^{2}_{I,i} \right\rbrace| \tau\right) \mathrm{e}^{\frac{1}{12}i\pi (\bm{N}^TK\bm{N}-\kappa )} \mathrm{e}^{i2\pi h_{\bm{\bm{\alpha}}}}
\end{eqnarray}
We have used some useful relations(here, we assume the total flux $N_{\phi}$ through torus is even):
\begin{eqnarray}\label{Modular-Properties}
\nonumber
\eta(\tau+1) &=& e^{i\pi/12}\eta(\tau)\\
\nonumber
\theta_{11}(z|\tau+1)    &=& e^{i\pi/4}\theta_{11}(z|\tau)\\
\nonumber
f^{(\bm{\alpha},0)}(\bm{Z}|\tau+1)  &=&   e^{i\pi \bm{\alpha}^TK\bm{\alpha}}f^{(\bm{\alpha},0)}(\bm{Z}|\tau) \\
f^{(\bm{\alpha},K^{-1}\bm{\kappa}/2)}(\bm{Z}|\tau+1)  &=&   e^{i\pi (\bm{\alpha}+\frac12 K^{-1}\bm{\kappa})^TK(\bm{\alpha}+\frac12 K^{-1}\bm{\kappa})}f^{(\bm{\alpha},K^{-1}\bm{\kappa}/2)}(\bm{Z}|\tau).
\end{eqnarray}
Finally, we derive the modular matrix representation:
\begin{equation}
\langle \Psi^{\bm{\beta}};\tau|\mathcal{T}|\Psi^{\bm{\alpha}};\tau\rangle = \delta_{\bm{\alpha}\bm{\beta}}  \mathrm{e}^{i2\pi (h_{\bm{\alpha}}-\frac{c}{24})} \mathrm{e}^{\frac{1}{12}i\pi \bm{N}^TK\bm{N}}.
\end{equation}
Here, $c=\kappa$ is the chiral central charge and $h_{\bm{\alpha}}$ is the topological spin of the topological sector $\bm{\alpha}$:
\begin{eqnarray}\label{topological-spin}
\nonumber 
\text{Boson:}~~~~h_{\bm{\alpha}}	&=&	\frac12 \bm{\alpha}^T K \bm{\alpha} \\
\text{Fermion:}~~~~h_{\bm{\alpha}}	&=&	\frac12 (\bm{\alpha}+ K^{-1}\bm{\kappa}/2)^T K (\bm{\alpha}+ K^{-1}\bm{\kappa}/2).
\end{eqnarray}

\subsection{Hall viscosity and Adiabatic phase}\label{hall-viscosity}
For convenience, we denote the wave function in Eq. \ref{Abelian-Wave} as $\frac{1}{\sqrt{Z(\tau,\bar{\tau})}}|\Phi;\tau\rangle $ with
\begin{eqnarray}
\langle \lbrace z^I_j \rbrace | \Phi;\tau\rangle &=& N_0 \eta(\tau)^{\bm{\kappa}^T\bm{N}} f_c(\left\lbrace Z^I \right\rbrace|\tau ) f_r( \left\lbrace z_i^I \right\rbrace|\tau ) 
\mathrm{e}^{i\pi N_{\phi} \tau \sum_{I,i}\left( y_i^I/L\tau_2\right)^2} \\
\nonumber
\langle \Phi;\tau | \Phi;\tau\rangle &=& \tau_2^{-\frac12 \bm{\kappa}^T\bm{N}} = \left( \frac{\tau-\bar{\tau}}{2i} \right) ^{-\frac12 \bm{\kappa}^T\bm{N}}=Z(\tau,\bar{\tau})
\end{eqnarray}
and 
$Z(\tau,\bar{\tau})=(\frac{\tau-\bar{\tau}}{2i} )^{-\frac12 \bm{\kappa}^T\bm{N}}$ is a pure real part. Thus, we can calculate the Berry connection:\cite{Read2009,Fremling2014}
\begin{eqnarray}\label{Berry-Connection-s}
\nonumber
\mathcal{A}_{\tau}&=&i\langle \Psi;\bar{\tau}|\frac{1}{\sqrt{Z}}\partial_{\tau}\frac{1}{\sqrt{Z}}|\Psi;\tau\rangle = 
i\sqrt{Z}\partial_{\tau}\frac{1}{\sqrt{Z}}+\frac{i}{Z}\partial_{\tau}Z=\frac i2 \partial_{\tau}\ln{Z}
=-\frac{\bm{\kappa}^T\bm{N}}{8\tau_2}\\
\mathcal{A}_{\bar{\tau}}&=&i\langle \Psi;\bar{\tau}|\frac{1}{\sqrt{Z}}\partial_{\bar{\tau}}\frac{1}{\sqrt{Z}}|\Psi;\tau\rangle = 
i\sqrt{Z}\partial_{\tau}\frac{1}{\sqrt{Z}}=-\frac i2 \partial_{\bar{\tau}}\ln{Z}
=-\frac{\bm{\kappa}^T\bm{N}}{8\tau_2}.
\end{eqnarray}
We can also rewrite Eq. \ref{Berry-Connection-s} in terms of $\mathcal{A}_{\tau_1}$ and $\mathcal{A}_{\tau_2}$:
\begin{eqnarray}\label{Berry-Connection12}
\nonumber
\mathcal{A}_{\tau_1}&=& \mathcal{A}_{\tau}+\mathcal{A}_{\bar{\tau}} = -\frac{\bm{\kappa}^T\bm{N}}{4\tau_2}\\
\mathcal{A}_{\tau_2}&=& i\mathcal{A}_{\tau}-i\mathcal{A}_{\bar{\tau}} = 0.
\end{eqnarray}
The $\mathcal{T}$ transformation induced adiabatical phase is:\cite{Fremling2014,Zaletel2013}
\begin{equation}\label{Berry-Phase-s}
\int_0^1\mathcal{A}_{\tau_1} \mathrm{d}\tau_1= -\frac{\bm{\kappa}^T\bm{N}}{4\tau_2}= -\hbar^{-1} \eta^H|L|^2.
\end{equation}
Finally, let us consider some examples. Laughlin state $\nu=1/q$, $\eta^H=\frac{\hbar qN_e}{4\tau_2 |L|^2}=\frac\hbar {N_{\phi}}{8\pi N_{\phi} l^2}
=\frac{\hbar 1/q}{8\pi l^2}q=\frac{\hbar \nu}{8\pi l^2}\mathcal{S}$ with topological shift $\mathcal{S}=q$. Halperin$(mmn)$ state, 
$\eta^H=\frac{\hbar mN_e}{8\pi N_{\phi} l^2}=\frac{\hbar \nu}{8\pi l^2}\mathcal{S}$ with topological shift $\mathcal{S}=m$.

\section{More numerical results}
\subsection{Laughlin state}
\begin{figure}[h]
	\includegraphics[width=1.0\textwidth]{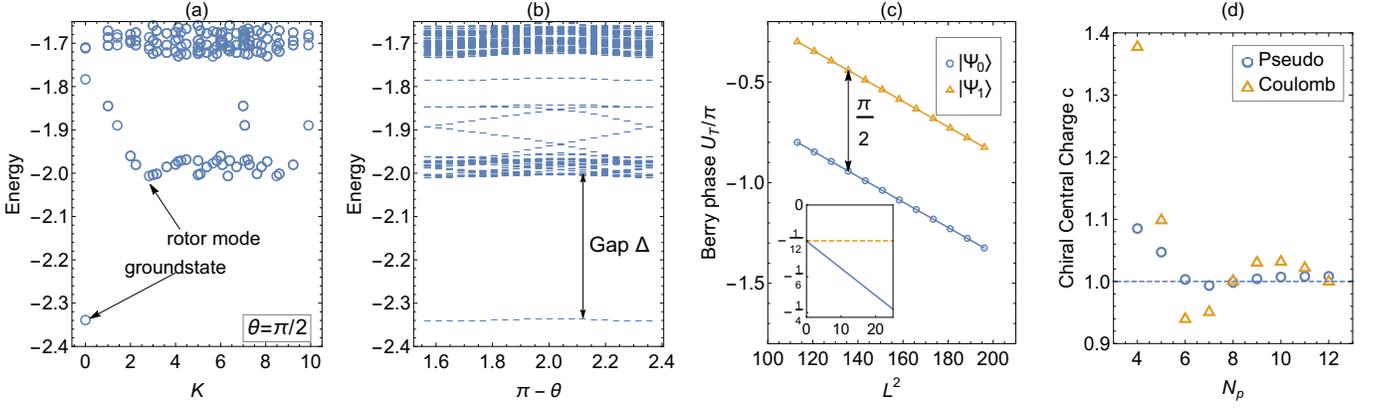}
	\caption{(a) Low-energy spectra of Laughlin $\nu=1/2$ state for geometric parameter $\theta=\pi/2$ and system size $N_p=8$.
		(b) Flow of energy spectra with varying geometric parameter $\theta$.
		The energy gap is defined by the minimal difference between ground states and excited states in the whole process of Dehn twist.
		(c) Accumulating Berry phase for different topological ground states $|\Psi_a\rangle$ with $a=0,1$ labeled by fractional quasiparticle charge of $Q=a/2$ (in unit of $e$).
		Through the linear fitting, the obtained guiding center spin is $s\approx-0.4997$ and topological spin is $h_1-h_0=0.2500$.The inset
		shows that the intercept of $h_{0}$ sector, therefore we obtain the chiral central charge $c\approx 0.9997$(the yellow dashed line is $-2\pi/24$). The system size is $N_p=12$.
		Here we choose Coulomb interaction and the geometric path in Fig. \ref{fig:torus}.
		(d) Chiral central charge $c$ with varying system size $N_p$, blue circles stand for pseudopotential and yellow triangles Coulomb interaction. The horizontal dashed line is $c=1$.
	} \label{sfig:laughlin-2}
\end{figure}

Fig. \ref{sfig:laughlin-2}(a) shows low-energy spectra versus momentum $K=\sqrt{K_1^2+K_2^2}$ of Laughlin $\nu=1/2$ state
by fixing geometric parameter $\theta=\pi/2$ by fixing $\tau_2=1$ (symmetric rectangular).
There is one single ground state in momentum $(0,0)$ 
, which is separated from the excited levels by a finite energy gap.
Considering the central mass degeneracy $q=2$ for $\nu=p/q=1/2$, we recover the two-fold ground-state degeneracy for Laughlin $\nu=1/2$ state.
Above the ground state, a magneto-roton branch exists $K>1$, representing the collective mode of quasielectron-quasihole pair.
In Fig. \ref{sfig:laughlin-2}(b),
varying the geometric parameter $\theta$ by fixing system area $|\vec L_1\times\vec L_2|$,
the ground state in momentum $(0,0)$ 
evolves adiabatically and never crosses with the higher energy levels in the spectral flow.
By collecting the accumulating phase in one deformation ($\tau \rightarrow \tau+1$) for different system parameter $L$, we get the
plot of $U_{\mathcal{T}}$ versus $L$, as shown in Fig. \ref{sfig:laughlin-2}.
Fitting by the linear relation Eq. \ref{eq:berryphase},
we get the guiding center spin is $s\approx-0.4997$ and topological spin is $h_1-h_0=0.2500$ (machine precision),
consistent with the prediction of Laughlin $1/2$ state with $s=-1/2$\cite{YeJePark2014} and $h_1-h_0=1/4$\cite{Wen2012}.
In particular, topological spin $h_1-h_0=1/4$ signals the element quasiparticles satisfy semionic statistics
that a semion goes back to itself by a self-rotation $8\pi$.

The same procedure can be applied to Laughlin $\nu=1/4$ state, as shown in Fig. \ref{sfig:laughlin-4}.
Comparison with $\nu=1/2$ case,  the only difference is there are four topologically distinct ground states at $\nu=1/4$.
We label the four different ground states by their fractional charge $Q=a/4$ (in unit of $e$, $e$ the element charge of electron), with $a=0,1,2,3$.
The different ground states can be distinguished by their topological spin: $h_{1,3}-h_0=0.1250$ and $h_2-h_0=0.5000$.
Combined the $\nu=1/2$ and $1/4$ results, we conclude that the quasiparticle $a$ in bosonic Laughlin $1/q$ state ($q$ even integer) carries topological spin $h_a-h_0=\frac{a^2}{2q}$.
This expression is consistent with Eq.\ref{topological-spin}.

\begin{figure}[h]
	\includegraphics[width=1.0\textwidth]{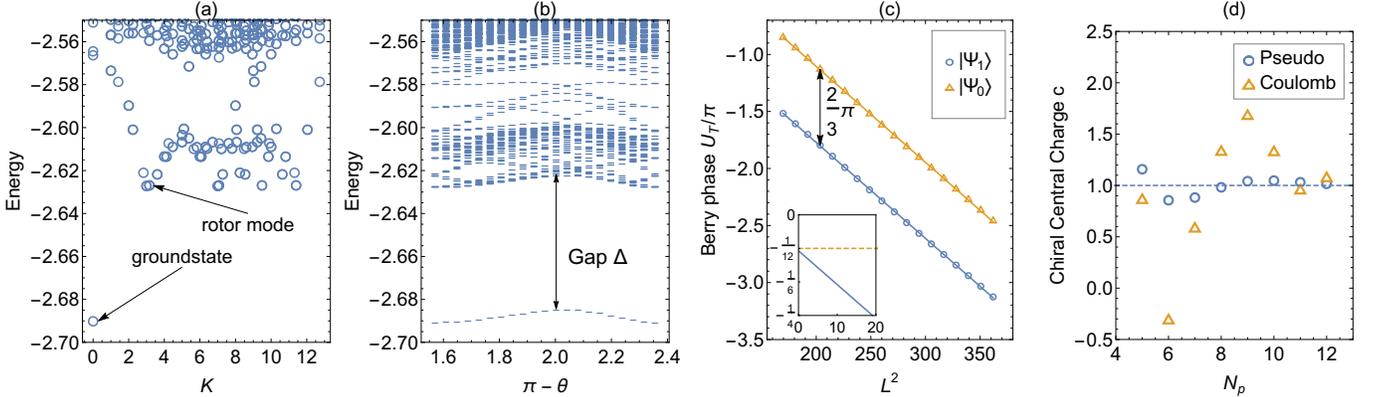}
	\caption{
		(a) Low-energy spectra for fermionic FQH $\nu=1/3$ state with system size $N_p=10$,
		by setting geometric parameter $\theta=\frac{\pi}{2}$  and $\tau_2=1$ (symmetric rectangular).
		The ground state is located in momentum sectors $(0,0)$.
		(b) Flow of energy spectra with varying geometric parameter $\theta$ with system size $N_p=10$.
		(c) Accumulating Berry phase for $N_p=12$. The different topological ground states $|\Psi_a\rangle$ with $a=0,1,2$ labeled by fractional quasiparticle charge of $Q=a/3$ (in unit of $e$). And $|\Psi_1\rangle$ and $|\Psi_2\rangle$ are equivalent.
		The obtained guiding center spin is $s\approx -0.9964$ and topological spin is $h_{0}-h_{1,2}\approx0.3333$.The inset
		shows that the intercept of $h_{1,2}$ sector, therefore we obtain the chiral central charge $c\approx 1.0699$(the yellow dashed line is $-2\pi/24$).
		Here we choose Coulomb interaction and the geometric path in Fig. \ref{fig:torus}.
		(d) Chiral central charge $c$ with varying system size $N_p$, blue circles stand for pseudopotential and yellow triangles Coulomb interaction. The horizontal dashed line is $c=1$.
	} \label{sfig:laughlin-3}
\end{figure}

\begin{figure}[h]
	\includegraphics[width=1.0\textwidth]{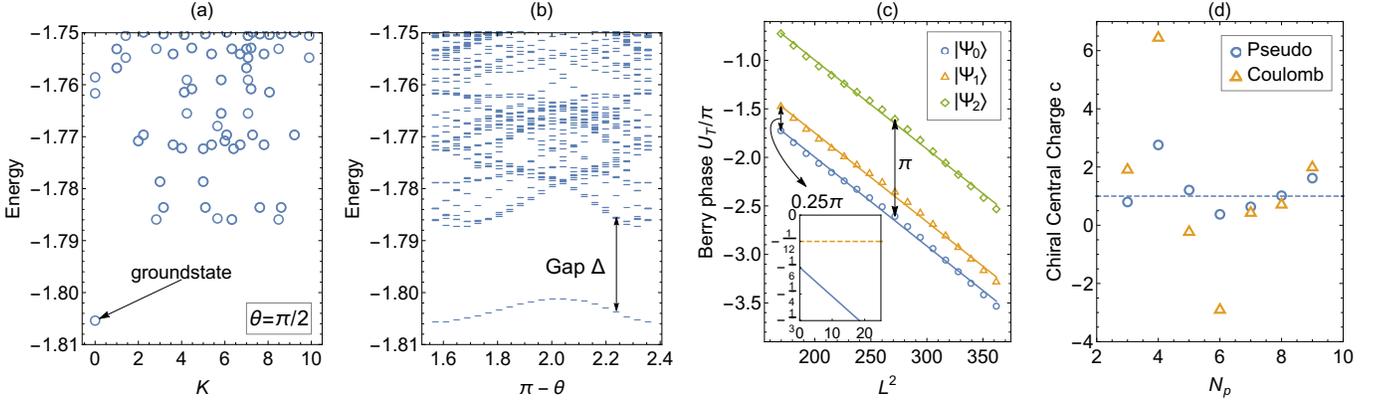}
	\caption{$\mathcal{T}-$transfermation on Laughlin state at $\nu=1/4$.
		(a) Low-energy spectra for $N_p=8$, by setting geometric parameter $\theta=\frac{\pi}{2}$  and $\tau_2=1$ (symmetric rectangular). The ground state is located in momentum sectors $(0,0)$.
		(b) Flow of energy spectra with varying geometric parameter $\theta$ for $N_p=8$.
		(c) Accumulating Berry phase for different topological ground states $|\Psi_a\rangle$ with $a=0,1,2,3$ labeled by fractional quasiparticle charge of $Q=a/4$ (in unit of $e$). And $|\Psi_1\rangle$ and $|\Psi_3\rangle$ are equivalent.
		The obtained guiding center spin is $s=-1.4469$ and topological spin is $h_{1,3}-h_0=0.1250,h_2-h_0=0.5000$. The inset
		shows that the central charge $c$ seems not converge for the largest system size $N_p=9$ that we can touch.
		Here we choose Coulomb interaction and the geometric path in Fig. \ref{fig:torus}.
		(d) Chiral central charge $c$ with varying system size $N_p$, blue circles stand for pseudopotential and yellow triangles Coulomb interaction. The horizontal dashed line is $c=1$.
	}\label{sfig:laughlin-4}
\end{figure}

For fermionic Laughlin $\nu=1/3$ state, we get very similar results, as shown in Fig. \ref{sfig:laughlin-3}.
The obtained guiding center spin $s\approx-0.9964$, which is very close to the theoretical prediction
from the Jack polynomial calculation\cite{YeJePark2014} and Laughlin wave function in Sec.~\ref{hall-viscosity}: $s=-1$.
Furthermore, based on the Jack polynomial formula in Ref. \cite{YeJePark2014}, the predicted topological spins
for three topological sector with $Q=a/3$ ($a=0,1,2$)  are {$h_1-h_0=h_2-h_0=1/6$}.
But in our calculation, we obtain the topological spin as $h_0-h_1=h_0-h_2\approx0.3333=1/3$ 
with machine precision.
The difference in topological spin has been noticed in Ref. \cite{Zaletel2013}:
The periodic boundary condition for fermion will lead to an additional $\pi$ phase so that the topological spin. The theoretical result in Eq.\ref{topological-spin} and Ref. \cite{Wen2012} is
$h_a=\frac{(a+q/2)^2}{2q}$, therefore we have $h_0=1/24+1/3=3/8,h_1=h_2=1/24$ which is consistent with numerical results .

\newpage
\subsection{Halperin state}
Here we discuss a class of FQH state for double-layer systems \cite{Eisenstein1992,Suen1992,WZhu2015,ZhaoLiu2015}  which called Halperin state. The Halperin states$(m,m,m-1)$ which filling factor is $\nu=2/(2m-1)=p/q$ and topological shift $\mathcal{S}=m$\cite{Wen1992}, thus we can derive 
the corresponding guiding center spin is $s=p(1-m)/2$.

Here we study fermionic $(332)$ and bosonic $(221)$ state. Considering $(332)$ state with $|\det{(K)}|=5$ degenerate ground states. The uint cell\cite{Fremling2014,Wen1993} includes 5 lattice points: 
$\left\lbrace \bm{\alpha} | (0,0),(1/5,1/5),(2/5,2/5),(3/5,3/5),(4/5,4/5)\right\rbrace $. Using Eq. \ref{topological-spin} we can calculate the theoretical topological spin 
for $(332)$ state are $h_{0,0}=h_{2,2}=9/20,h_{3,3}=h_{4,4}=1/20,h_{1}=1/4$. 
For bosonic $(221)$ state, the uint cell has $|\det{(K)}|=3$ lattice points:
$\left\lbrace \bm{\alpha} | (0,0),(1/3,1/3),(2/3,2/3)\right\rbrace $ and the corresponding topological spin are $h_{0,0}=0,h_{1,1}=h_{2,2}=1/3$.

\begin{figure}[h]
	\includegraphics[width=1.0\textwidth]{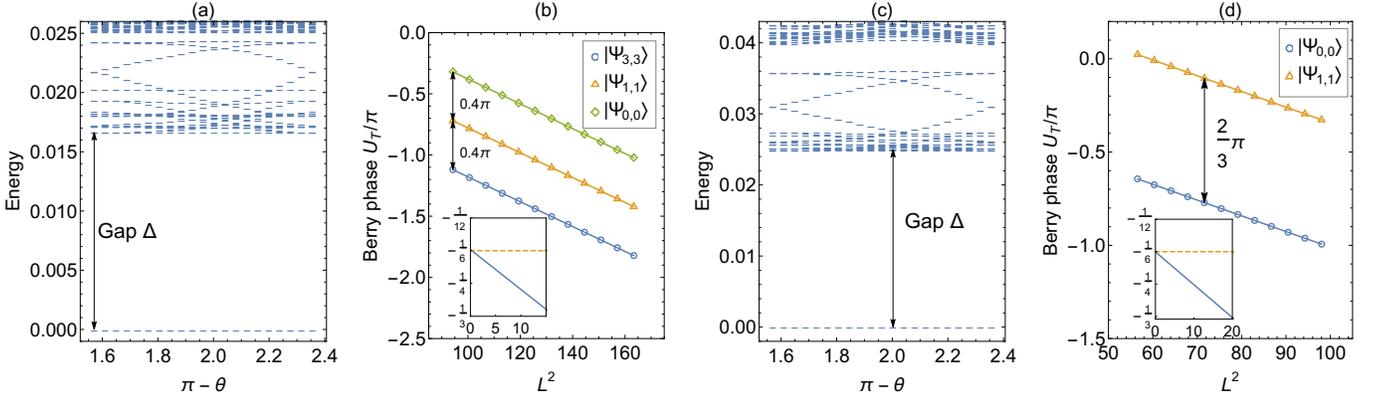}
	\caption{
		(a-b)$\mathcal{T}-$transfermation on Halperin$(332)$ state for $N_p=8$, we choose Pseudopotential and the geometric path in Fig. \ref{fig:torus}.
		(a) Flow of energy spectra with varying geometric parameter $\theta$.
		(b) Accumulating Berry phase for different topological ground states $ |\Psi_{a,a} \rangle$ with $a=0,1,3$.
		The obtained guiding center spin is $s\approx-2.0033$ and topological spin is $h_{0,0}-h_{1,1}=h_{2,2}-h_{1,1}=0.2000,h_{1,1}-h_{3,3}=h_{1,1}-h_{4,4}=0.2000$.
		The inset shows that the intercept of $h_{3,3}$ sector, therefore we obtain the chiral central charge $c\approx 1.9678$(the yellow dashed line is $-4\pi/24$).
		(c-d)$\mathcal{T}-$transformation on Halperin$(221)$ state for $N_p=8$, we choose pseudopotential and the geometric path in Fig. \ref{fig:torus}.
		(c) Flow of energy spectra with varying geometric parameter $\theta$.
		(d) Accumulating Berry phase for different topological ground states $ |\Psi_{a,a} \rangle$ with $a=0,1,3$.
		The obtained guiding center spin is $s\approx-0.9986$ and topological spin is $h_{1,1}-h_{0,0}=h_{2,2}-h_{0,0}=0.3333$.
		The inset shows that the intercept of $h_{0,0}$ sector, therefore we obtain the chiral central charge $c\approx 2.0080$(the yellow dashed line is $-4\pi/24$).
	}
\end{figure}

\newpage
\subsection{Particle-Hole Conjugate $\nu=2/3$ State}
The fermionic $\nu=2/3$ state is the particle-hole conjugate state of Laughlin $\nu=1/3$ state, whose $K$ matrix is\cite{Wen_book}:
\begin{eqnarray}
K=\left(
\begin{array}
{cc}
1	&	1\\
1		&	-2
\end{array}
\right)
\end{eqnarray}
and ground states are also 3-fold degenerated. Using the Eq. (12) in Ref. \cite{Wen1992}, we can derive the topological shift $\mathcal{S}=0$, therefore the guiding center spin 
is $s=p(\tilde{s} - \mathcal{S}/2)=1$ and the topological spin is $h_0=-1/4,h_1=h_2=1/12$. 

\begin{figure}[h]
	\includegraphics[width=1\textwidth]{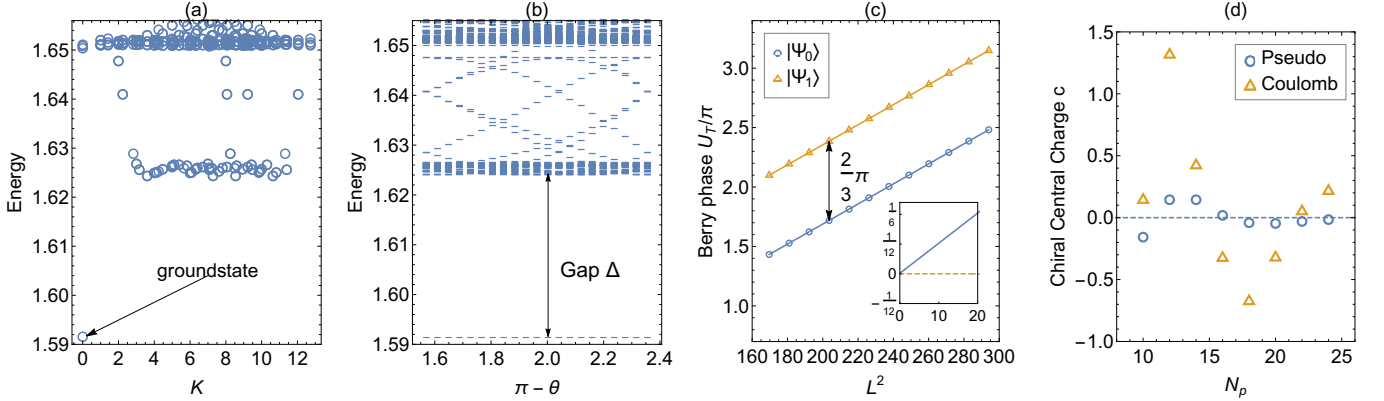}
	\caption{(a) Low-energy spectra for fermionic FQH $\nu=2/3$ state with system size $N_p=20$,
		by setting geometric parameter $\theta=\frac{\pi}{2}$  and $\tau_2=1$ (symmetric rectangular).
		The ground state is located in momentum sectors $(0,0)$.
		(b) Flow of energy spectra with varying geometric parameter $\theta$ with system size $N_p=20$.
		(c) Accumulating Berry phase for $N_p=24$. 
		The obtained guiding center spin is $s\approx -0.9990$ and topological spin is $h_{1,2}-h_{0}\approx0.3333$.The inset
		shows that the intercept of $h_{0}$ sector, therefore we obtain the chiral central charge $c\approx 0.0160$(the yellow dashed line is $0$).
		Here we choose $v_1=1$ pseudopotential and the geometric path in Fig. \ref{fig:torus}.
		(d) Chiral central charge $c$ with varying system size $N_p$, blue circles stand for $v_1=1$ pseudopotential and yellow triangles Coulomb interaction. The horizontal dashed line is $c=0$.
	}
\end{figure}

\subsection{Hierarchy state}
Starting from Laughlin $\nu= 1/q$ state, quasiparticles can condensate into successive Laughlin states
and generate a hierarchy of incompressible states. In the case of fermions the most prominent series are given by $\nu=\frac{2}{5}$,
while for bosons $\nu=\frac{2}{3}$.
And the relationship between the topological shift and filling factor is $\mathcal{S}=4$ for fermions and $\mathcal{S}=3$ for bosons.
Thus, we anticipate that the guiding center spin is $s=-3$ for fermions $\nu=\frac{2}{5}$ and $s=-2$ for bosons $\nu=\frac{2}{3}$.

Our numerical simulation gives the guiding center spin $s\approx-2.0840$ for the bosonic $\nu=2/3$ state, and $s\approx-2.9552$ for  the fermionic $\nu=2/5$ state, both of which matches the above expected values.
Moreover, we also estimate the topological spin of element quasiparticle in $\nu=2/3$ is $h_{1,2}-h_0\approx0.3333=1/3$
and in $2/5$ is $h_{1,4}-h_0\approx0.2=1/5,h_{2,3}-h_0\approx0.4=2/5$.

\begin{figure}[h]
	\includegraphics[width=1\textwidth]{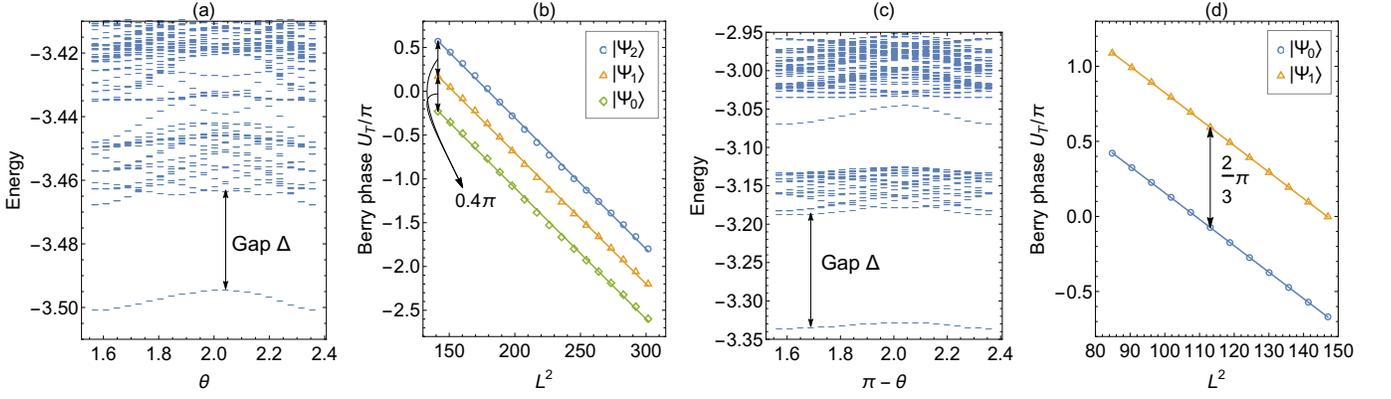}
	\caption{$\mathcal{T}-$transformation on fermion Hierarchy state for (a)-(b) fermionic $\nu=2/5$ and (c)-(d) bosonic $\nu=2/3$ with system size $N_p=12$. 
		Here we choose the geometric path in Fig. \ref{fig:torus}.
		(a) Flow of energy spectra with varying geometric parameter $\theta$.
		(b) Accumulating Berry phase for different topological ground states $|\Psi_a\rangle$ with $a=0,1,2,3,4$.
		The obtained guiding center spin is $s\approx-2.9552$ and topological spin is $h_{1,4}-h_0=0.2000,h_{2,3}-h_0=0.4000$.
		(c) Flow of energy spectra with varying geometric parameter $\theta$.
		(d) Accumulating Berry phase for different topological ground states $|\Psi_a\rangle$ with $a=0,1,2$.
		The obtained guiding center spin is $s\approx-2.0840$ and topological spin is $h_{1,2}-h_0=0.3333$.
	}
\end{figure}

\subsection{Fermionic Moore-Read state}

\begin{figure}[h]
	\includegraphics[width=0.75\textwidth]{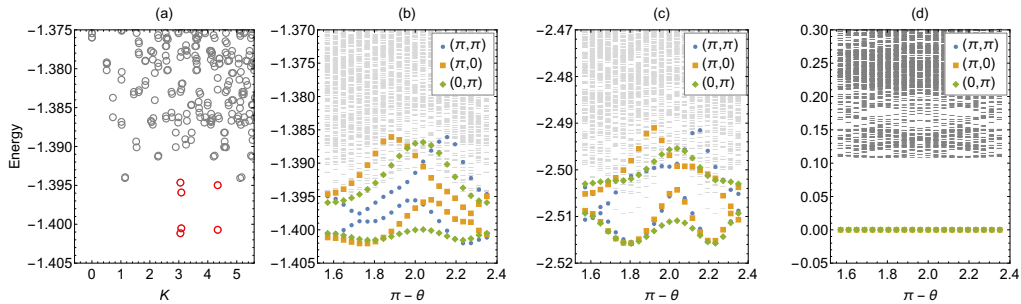}
	\caption{(a) Low-energy spectra for fermionic FQH $\nu=5/2$ state with system size $N_p=12$ by setting
		$\theta=\frac{\pi}{2}$ and $\tau_2^{-1}=0.99$ \cite{Peterson2008}.
		The model contains a finite-layer width correction to the pure Coulomb interaction.
		The six-fold quasi-degenerated ground states in momentum sectors $(\pi,0),(0,\pi),(\pi,\pi)$ are indicated by red circle.
		(b) Flow of low-energy spectra with changing torus geometry $\theta$.
		The ground states in momentum $(\pi,\pi),(\pi,0),(0,\pi)$ are labeled by blue circle, yellow square and green rhombus.
		(c) Low-energy spectra for fermionic FQH $\nu=5/2$ state for pure Coulomb interaction. 
		(d) Low-energy spectra for fermionic FQH $\nu=5/2$ state for three-body model Hamiltonian.
	}\label{sfig:moore}
\end{figure}

As well-known that $\nu=p/q$ (with $q$ odd) can be understood by
the Laughlin paradigm and further hierarchy theory or Jain composite fermion theory,
the finding of even denominator $\nu=5/2$ FQH state challenges
our theoretical understanding of the FQH effect.
Among all candidates, Pfaffian or anti-Pfaffian wave function proposed by
Moore and Read \cite{Greiter1991,Moore1991,Levin2007} seems a  promising candidate to describe the enigmatic nature of FQH $\nu=5/2$ state.
Although much efforts have been devoted to this long-standing issue  \cite{Morf1998,Haldane2000,Morf2010,Pakrouski2015,Peterson2008,HaoWang2009},
solid numerical evidence of topological ground state degeneracy on torus is still lacking.
In the main text, we have shown that the quasi-degenerate ground state of pure Coulomb interaction
are not stable against the dehn twist transformation on the limited system size.
We also noticed that in another study, Ref. \cite{Peterson2008} proposed that
the modified Coulomb interaction with finite-layer width correction
may enhance the Moore-Read signature in some range of aspect ratio on torus.
Here we also investigate this possibility, by using the modified Coulomb potential in
infinite square-well potential:
\begin{equation}
V(k)=\frac{2\pi}{k}\frac{3kd+\frac{8\pi^2}{kd}-\frac{32\pi^4(1-e^{-kd})}{k^2d^2(k^2d^2+4\pi^2)}}{k^2d^2+4\pi^2}
\end{equation}
where $d$ stands for the effective layer-width of the experimental GaAs quantum well structures.
In the calculation, we set $d=4 \ell$ according to the discussion in Ref. \cite{Peterson2008}.
The low-energy spectra at rectangular geometry is shown in Fig.\ref{sfig:moore}(a),
which should exactly repeat the results of Fig. 4 in Ref. \cite{Peterson2008}.
The plausible six-fold quasi-degenerate ground states are labeled by red circle.
However, under the Dehn twist deformation, the six-fold quasi-degenerate states
evolve into the higher levels, as shown in Fig. \ref{sfig:moore}(b).
Due to the fail of parallel transport, we cannot get Hall viscosity and topological spin for $\nu=5/2$ state
for Coulomb interaction in our calculation.
Here, our analysis based on geometric deformation suggests that
numerical signature of Moore-Read state on torus geometry is still questionable.

How to understand our results on $\nu=5/2$ quantum Hall state?
One possible understanding is that,
the Coulomb ground states at $\nu=5/2$ lie on the marginal boundary between Pfaffian and anti-Pfaffian state
since particle-hole symmetry cannot be broken on torus geometry \cite{HaoWang2009}.
The recent progresses of discovering non-Abelian statistics of FQH $\nu=\frac{5}{2}$ state on
cylinder and sphere geometry may shed some light on this issue, where
the particle-hole symmetry is broken spontaneously or explicitly \cite{Haldane2008,Zaletel2015,Pakrouski2015}.
In addition, recent thermal Hall measurement brings other possibility to our attention. 
For example, the particle-hole preserved Pfaffian state is proposed as a viable possibility \cite{Feldman2016}. 
Particle-hole Pfaffian state should host three-fold ground state degeneracy (excluding the center-of-mass degeneracy).
Unfortunately, in our extensive calculations (see Fig. \ref{sfig:moore}), 
we did not observe signal for three-fold ground state degeneracy neither. 
In a word, our results call for further study on the $\nu=5/2$ problem on the torus geometry.


\subsection{Bosonic Moore-Read State}
\begin{figure}[h]
	\includegraphics[width=0.75\textwidth]{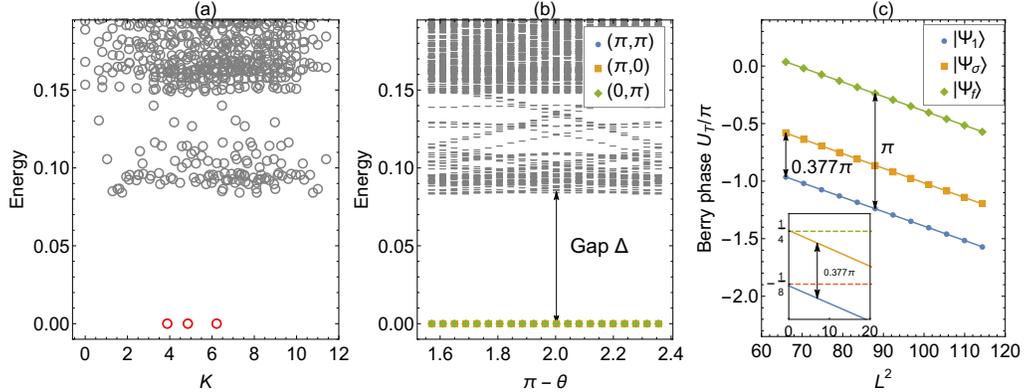}
	\caption{(a) Low-energy spectra for bosonic Moore-Read $\nu=1$ state with system size $N_p=12$,
		by setting geometric parameter $\theta=\frac{\pi}{2}$  and $\tau_2=1.25$.
		The three-fold degenerated ground states in momentum sectors $(\pi,0),(0,\pi),(\pi,\pi)$ are indicated by red circle.
		(b) Flow of low-energy spectra with changing torus geometry $\theta$.
		The ground states in momentum $(\pi,\pi),(\pi,0),(0,\pi)$ are labeled by blue circle, yellow square and green rhombus.
		(c) Accumulating Berry phase for $N_p=14$. 
		The obtained guiding center spin is $s\approx -1.0039(-0.9898)$ for $|\Psi_\sigma\rangle(|\Psi_1\rangle)$ and topological spin is $h_{f}-h_{1}\approx0.5000,h_{\sigma}-h_{1}\approx0.1885$. The inset
		shows that the intercept of $h_{1}$ and $h_\sigma$ sector, therefore we obtain the chiral central charge $c\approx 1.4276$(the green dashed line is $-\frac18+\frac38=\frac14$) for $h_\sigma$ and 
		$c\approx 1.6441$(the blue dashed line is $-\frac18$) for $h_1$, the average of these two sectors is $c\approx 1.5359$.
		Here we choose the geometric path in Fig. \ref{fig:torus}.
	}\label{boson-MR-model}
\end{figure}

In the main text, the bosonic Moore-Read state is studied based on Coulomb potential.
Here, we investigate the bosonic Moore-Read $\nu=1$ state using a 3-body model Hamiltonian:
\begin{eqnarray}\label{3body-boson}
V_{ijk}&=&\sum_{j_1,j_2,j_3,j_4,j_5,j_6} A_{j_6,j_5,j_4}^{j_1,j_2,j_3} 
a_{j_1}^\dagger a_{j_2}^\dagger a_{j_3}^\dagger a_{j_4} a_{j_5} a_{j_6}
\end{eqnarray}
where $A_{j_6,j_5,j_4}^{j_1,j_2,j_3}$ is:
\begin{eqnarray}
\nonumber
A_{j_6,j_5,j_4}^{j_1,j_2,j_3} 
&=& \frac{1}{\tau_2^2 L^4} \sum_{\bm{q}_1,\bm{q}_2} \exp{\left\lbrace  -\frac14 \left[ \bm{q}_1^2+(\bm{q}_1-\bm{q}_2)^2+\bm{q}_2^2\right] \ell^2\right\rbrace }\delta^{\mathrm{mod} N_{\phi}}_{j_1+j_2+j_3,j_4+j_5+j_6}
\\ \nonumber &&\times 
\left\lbrace \exp{\left[  - i2\pi \frac{n_{1y}}{N_\phi} (j_6-j_2+\frac{n_{1x}}{2})-i2\pi \frac{n_{2y}}{N_\phi} (j_2-j_4+\frac{n_{2x}}{2})\right]} \delta^{\mathrm{mod} N_{\phi}}_{j_1-j_6,n_{1x}}\delta^{\mathrm{mod} N_{\phi}}_{j_4-j_3,n_{2x}} \right.\\ \nonumber
&&+\exp{\left[ - i2\pi \frac{n_{1y}}{N_\phi} (j_5-j_3+\frac{n_{1x}}{2})-i2\pi \frac{n_{2y}}{N_\phi} (j_3-j_6+\frac{n_{2x}}{2})\right]} \delta^{\mathrm{mod} N_{\phi}}_{j_2-j_5,n_{1x}}\delta^{\mathrm{mod} N_{\phi}}_{j_6-j_1,n_{2x}} \\ 
&&\left.+ \exp{\left[\lbrace  - i2\pi \frac{n_{1y}}{N_\phi} (j_4-j_1+\frac{n_{1x}}{2})-i2\pi \frac{n_{2y}}{N_\phi} (j_1-j_5+\frac{n_{2x}}{2})\right] } \delta^{\mathrm{mod} N_{\phi}}_{j_3-j_4,n_{1x}}\delta^{\mathrm{mod} N_{\phi}}_{j_5-j_2,n_{2x}}
\right\rbrace .
\end{eqnarray}
As shown in Fig. \ref{boson-MR-model}(a),
the three-fold degenerated ground states of bosonic Moore-Read $\nu=1$ state are the zero-energy state. 
According to the topological theory, these three states corresponding to topological sectors: two abelian anyon with topological spin $h_1=0,h_f=\frac12$ and a non-abelian Ising anyon $h_\sigma=\frac{3}{16}=0.375$.
The topological shift of Moore-Read $\nu=\frac pq=\frac 22=1$ state is $\mathcal{S}=2$ and guiding center spin $s=p(\tilde{s}-\frac{\mathcal{S}}{2})=-1$.
As shown in Fig. \ref{boson-MR-model}(c), our calculation gives guiding center spin is $s\approx -1.00$, which is consistent with the effective theory. 
And the topological spin also matches the theoretical prediction.
In addition, the average chiral central charge is determined to be $c\approx 1.5359$, close to the theoretical prediction $c=3/2$. 
These facts form a compete diagnosis of non-Abelian nature of bosonic Moore-Read state at $\nu=1$.

\clearpage
\section{Geometric Path Dependence}
\begin{figure}[h]
	\includegraphics[width=0.65\linewidth]{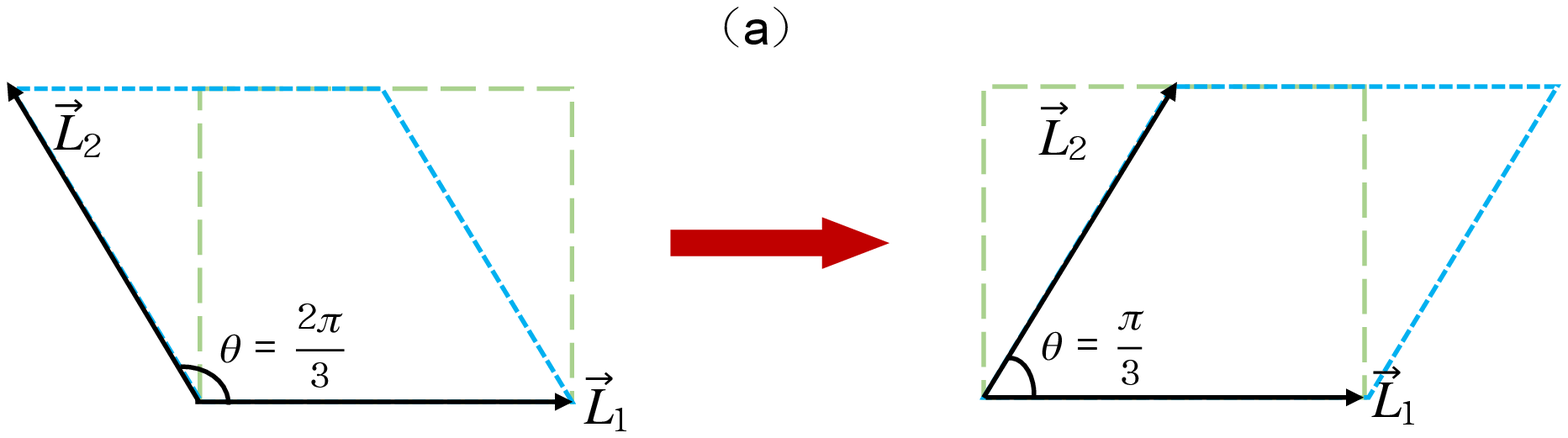}
	\includegraphics[width=0.75\textwidth]{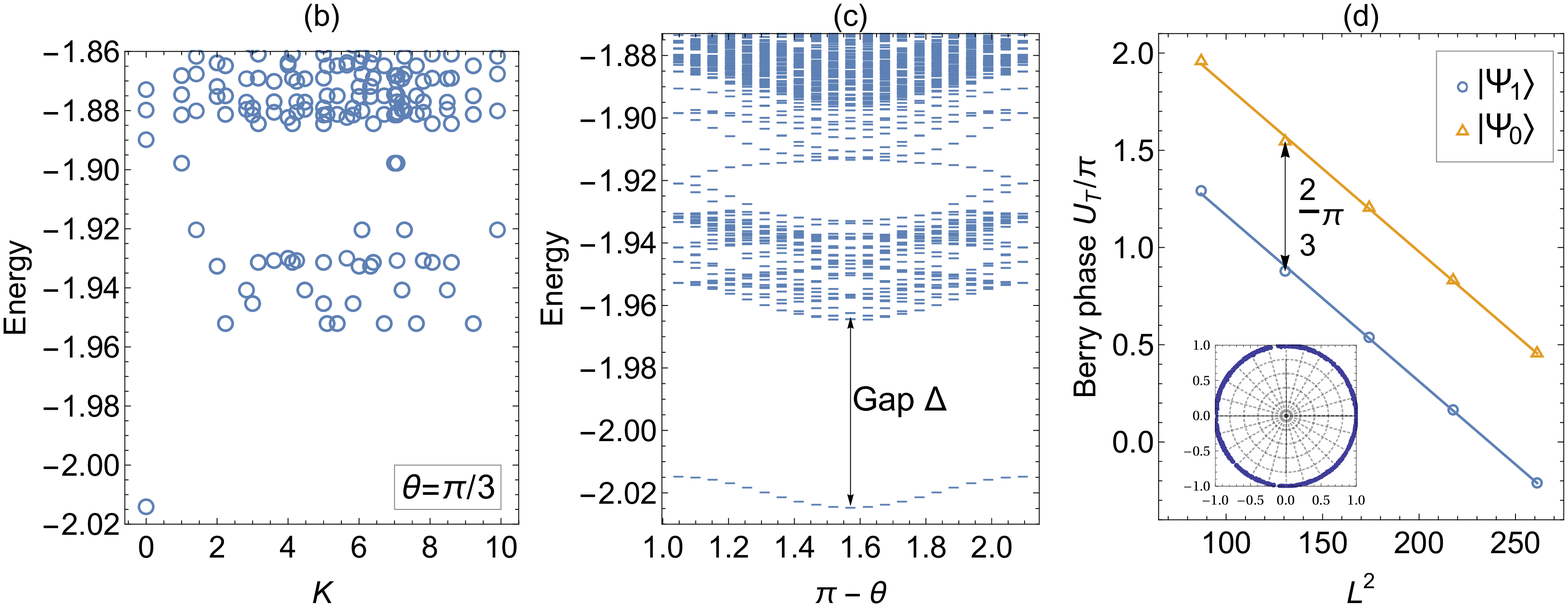}
	\caption{(a) The geometric path we used here: $\theta: 2\pi/3\rightarrow \pi/3$. (b) Low-energy spectra for fermionic FQH $\nu=1/3$ state with system size $N_p=8$,
		by setting symmetrical hexagonal geometry ($\theta=\frac{\pi}{3}$ and $\tau_2=1$).
		The ground state is located in momentum sectors $(0,0)$.
		(c) Flow of energy spectra with varying geometric parameter $\theta$.
		(d) Accumulating Berry phase for different topological ground states $|\Psi_a\rangle$, where $|\Psi_a\rangle$ labeled by fractional quasiparticle charge of $Q=a/3, a=0,1,2$ (in unit of $e$). And $|\Psi_1\rangle$ and $|\Psi_2\rangle$ are equivalent.
		The obtained guiding center spin is $s\approx -1.0119$ and topological spin is $h_{1,2}-h_0\approx0.3333$.
	} \label{sfig:hex}
\end{figure}

The geometric Berry phase is intrinsic under geometric deformation, which should not depend on the specific deformation path we choose.
To elucidate the physics does not depend on the specific geometric path we choose, we double check the different geometric paths in Fig. \ref{sfig:hex}(a): One is geometry path from a hexagon-like geometry to its equivalent one: $\theta: 2\pi/3 \rightarrow \pi/3$ (labeled by blue) and the other one is from rectangular to its equivalent one (labeled by red). For both geometric path, the extracted physics, guiding-center spin and topological spin are all identical.

Moreover, since the Berry phase only depends on the path length ($L$) along the $\vec L_1$ direction,  there are two different schemes to extract the Berry phase. One is to tune the aspect ratio $|\tau|=|\vec{L}_2|/|\vec{L}_1|$ and fix total flux $N_{\phi}$ of the system. The other one is to change the total flux of the system.
The topological information from these two different schemes should  coincide. As an example, in Fig. \ref{sfig:hex}(d), we show the Berry phase $U_\mathcal{T}$ for
system sizes $N_p=4,6,8,10,12$ by fixing the aspect ratio $|\tau|=1$(at the starting point of the $\mathcal{T}$ transformation).
In Fig. \ref{sfig:hex}(b), it is shown the low-energy spectra at symmetric hexagon geometry. Again, the ground state is located in momentum sector $(0,0)$.
The three-fold ground-state degeneracy is recovered when considering central-mass degeneracy of the system.
And we also observe a branch of magneto-roton modes above the ground state.
In Fig.\ref{sfig:hex}(c), when changing the geometric parameter $\theta$, the energy gap keeps open and the ground state indeed evolves adiabatically. In Fig. \ref{sfig:hex}(d), the Berry phase can be fitted by linear function. The topological information, guiding center spin $s\approx-1.0119\approx-1$ and topological spin $h_{1,2}-h_0\approx 0.3333\approx 1/3$, are consistent with the previous conclusion obtained by different geometric path.
Here, it is shown that the physics are identical when choosing different deformation paths.
Thus, reaching the same physics by the two different schemes is another evidence that the Berry phase under $\mathcal{T}-$transformation only depends on $L$, not on the total area $|\vec L_1\times \vec L_2|$ or total flux $N_{\phi}$.

%

\newpage
\section{Relation to other works} \label{sec:relation}
We recover the different definitions discussed in other works \cite{Wen1992,Read2009}. In our previous calculations, only the guiding center Hall viscosity has been considered since we are working on the projected Hamiltonian Eq. \ref{eq:projected-Ham}. Indeed, the total Hall viscosity of a FQH system includes two parts, guiding center Hall viscosity $\eta^g$ and Landau orbital Hall viscosity $\eta^o$.
The guiding-center Hall viscosity $\eta^g$ describes an emergent geometric response of the correlated
electrons, while Landau-orbit Hall viscosity $\eta^o$ directly comes from the Landau-orbit form factor.
For the Landau-orbit part, the Landau-orbit Hall viscosity can be expressed in terms
of Landau-orbit spin $\tilde{s}$ \cite{Avron1995}:
\begin{equation}\label{}
\eta^o=\frac{\hbar}{4\pi l^2} \nu \tilde{s},
\end{equation}
where Landau-orbit spin is defined as $\tilde{s}=n+\frac{1}{2}$ for $n-$th Landau level.
$\tilde{s}$ describes that, as the Landau level index increases,
the orbital angular momentum carried by the cyclotron motion also increases.
Thus, for a given filling factor $\nu$, Landau-orbit spin is a constant therefore Landau-orbit Hall viscosity is also a constant term.
Please note this Landau-orbit Hall viscosity
exists even when the particles are uncorrelated.

Combining the the Landau-orbit and guiding-center Hall viscosities, we reach the total Hall viscosity
\begin{equation}\label{}
\eta^H= \eta^g+\eta^o=\frac{\hbar}{4\pi l^2} (\nu \tilde{s}- \frac{s}{q})= \frac{\hbar}{4} \frac{\nu}{2\pi l^2} (2\overline{s}) = \frac{\hbar}{4} \frac{\nu}{2\pi l^2} \mathcal{S}
\end{equation}
Here we recover the so-called mean ``orbital spin'' defined by $\overline{s}=\tilde{s}-\frac{s}{p}$, which was first derived by Wen and Zee,\cite{Wen1992}
and later by Read and Rezayi\cite{Read2009}. And the relationship between the orbital spin $\overline{s}$ and the topological shift number $\mathcal{S}$ is \cite{Read2009}
\begin{equation}\label{}
\mathcal{S}=2\overline{s}.
\end{equation}
Topological shift $\mathcal{S}$ is a topological number related to a given FQH state, which
always vanishes on torus or plane geometry, but takes nonzero values on the curved geometry such as sphere.
Physically, the shift on curved space results from the nature of orbital spin carried by composite bosons in FQH effect.
In phenomenological way, the basic element in FQH effect at filling factor $\nu=p/q$ is a composite boson with
$p$ particles in $q$ consecutive orbitals.
Because composite bosons carry a non-zero orbital spin $\overline{s}$, the total flux seen by the composite boson is the sum of the magnetic flux and the Berry
phase induced through the coupling of the orbital spin $\overline{s}$ to the curvature of the space.
It is this Berry phase that causes the topological shift $\mathcal{S}$.
Thus, the composite bosons with different $\overline{s}$ will have different shifts $\mathcal{S}$ on sphere.
Here, we see these topological quantum numbers, $\mathcal{S}$ and the corresponding guiding center spin $s$,
can be measured by geometric twist, although they do not directly appear on the torus or plane.

We will close this section by illustrate some examples. For $\nu=p/q=1/q$ Laughlin states, the guiding center spin is $s=\frac{1}{2}(1-q)$.
The orbital spin per composite boson is $\overline{s}=\tilde{s}-s=\frac{q}{2}$, where we choose $\tilde{s}=1/2$ for lowest Landau level.
Therefore, topological shift $\mathcal{S}=2\overline{s}=q$.
The same procedure can be easily adapted to obtain the topological shift $\mathcal{S}=4$ for Hierarchy $\nu=2/5$ state.
These results coincide with prediction from by Wen and Zee,\cite{Wen1992} and Read
and Rezayi\cite{Read2009}, and previous works on spherical calculations.


\clearpage
\section{Minimal Entangled State in Bosonic $\nu=1$ case}

We give a symmetry analysis of bosonic Moore-Read $\nu=p/q=2/2=1$ state based on the root configuration.
Due to the condition of $N_p=N_{\phi}$, the center of mass degeneracy is $1$.
The root configuration for bosonic Moore-Read state is
no more than $p$ bosons in consecutive $q$ orbitals, thus we have three topological different ground states:
\begin{equation*}
\,\,\,\,\,[20], \,\,\,\,\,\,[02], \,\,\,\,\,\,\,[11]
\end{equation*}
These root configuration state also correspond to the ground states with definite type of anyonic quasiparticle,
which is the quasiparticle eigenstates.
These three ground states have different momentum quantum numbers along $K_x$: $[20]$ and $[02]$ have $K_x=1\%2$ (or $\pi$) while $[11]$ has $K_x=0\%2$ (or $0$).


Following the analysis in Ref. \cite{Bernevig2012}, we get the relationship between quasiparticle eigenstates 
and numerical obtained ground states labeled by momentum quantum number $(K_x,K_y)$ as :
\begin{eqnarray}\label{eq:mes}
|k_x=\pi,k_y=\pi> &=& |20 \rangle + e^{2\pi i 0/N} |02\rangle = |20 \rangle +  |02\rangle \nonumber\\
|k_x=\pi,k_y=0> &=& |20 \rangle + e^{2\pi i 1/N} |02\rangle = |20 \rangle -  |02\rangle \nonumber\\
|k_x=0,k_y=\pi> &=& |11 \rangle
\end{eqnarray}

To see Eq. \ref{eq:mes} really represents the quasiparticle eigenstates, we provide two different proves here.
One is that, we can  construct the modular $\mathcal{S}$ matrix based on Eq. \ref{eq:mes}, and the other one is, we can numerically prove that
Eq. \ref{eq:mes} are the minimal entangled states, which should be a faithful representation of quasiparticle eigenstates.

First, Eq. \ref{eq:mes} is the quasiparticle eigenstates, which are defined by coordinates $x$ and $y$:
\begin{eqnarray*}
	|\Xi^x_1\rangle&=&\frac{1}{\sqrt{2}}(|k_x=\pi,k_y=\pi \rangle +|k_x=\pi,k_y=0 \rangle)= |02 \rangle \nonumber\\
	|\Xi^x_2\rangle&=&\frac{1}{\sqrt{2}}(|k_x=\pi,k_y=\pi \rangle -|k_x=\pi,k_y=0 \rangle)= |20 \rangle \nonumber\\
	|\Xi^x_3\rangle&=&|k_x=0,k_y=\pi \rangle= |11\rangle
\end{eqnarray*}
Under the $\mathcal{S}$ transfermation, coordinates change to $x\rightarrow y$ and $y\rightarrow -x$, and we get the quasiparticle eigenstate as:
\begin{eqnarray*}
	|\Xi^y_1\rangle&=&\frac{1}{\sqrt{2}}(|k_x=\pi,k_y=\pi  \rangle +|k_y=\pi ,k_x=0\rangle)=\frac{1}{\sqrt{2}}[ \frac{1}{\sqrt{2}}(|02 \rangle +  |20\rangle)+ |11\rangle]\\
	|\Xi^y_2\rangle&=&\frac{1}{\sqrt{2}}(|k_x=\pi,k_y=\pi  \rangle -|k_y=\pi ,k_x=0\rangle)=\frac{1}{\sqrt{2}}[ \frac{1}{\sqrt{2}}(|02 \rangle +  |20\rangle)- |11\rangle]  \\
	|\Xi^y_3\rangle&=&|k_x=\pi,k_y=0 \rangle =\frac{1}{\sqrt{2}}[|02 \rangle -  |20 \rangle],
\end{eqnarray*}
since we applied a $\pi/2$ rotation on ground states:
\begin{eqnarray*}
	|k_x=\pi,k_y=\pi  \rangle\rightarrow |k_x=\pi,k_y=\pi   \rangle  \\
	|k_x=0,k_y=\pi \rangle\rightarrow |k_x=\pi, k_y=0 \rangle  \\
	|k_x=\pi,k_y=0 \rangle\rightarrow |k_x=0, k_y=\pi \rangle  \\
\end{eqnarray*}

Finally, we have,
\begin{equation*}
\mathcal{S}=\langle \Xi^x_i|\Xi^y_j\rangle=
\frac{1}{2}
\left(
\begin{array}{ccc}
1 & 1 & \sqrt{2} \\
1 & 1 & -\sqrt{2} \\
\sqrt{2} & -\sqrt{2} & 0
\end{array}
\right)
\end{equation*}
Therefore, since Eq. \ref{eq:mes} faithfully recovers the $\mathcal{S}$ matrix,
Eq. \ref{eq:mes} gives the quasiparticle eigenstates.

In practice, for twist transformation we denote the ground state by $|k_x,k_y,\tau(s)\rangle$, and let $\tau(s)=\tau(0)+s$. 
Since the numerical calculation adds an additional phase to each wave function, we must be careful to use Eq. \ref{eq:mes}. 
The quasiparticle state, or minimal entangled state can be written as 
\begin{eqnarray*}
	|\Xi^x_1,\tau\rangle&=&\frac{1}{\sqrt{2}}(|k_x=\pi ,k_y=\pi ,\tau\rangle +e^{i\varphi}|k_x=\pi,k_y=0,\tau\rangle) \nonumber\\
	|\Xi^x_2,\tau\rangle&=&\frac{1}{\sqrt{2}}(|k_x=\pi ,k_y=\pi ,\tau\rangle -e^{i\varphi}|k_x=\pi,k_y=0,\tau\rangle) \nonumber\\
	|\Xi^x_3,\tau\rangle&=&|k_x=0,k_y=\pi,\tau\rangle,
\end{eqnarray*}
where  $\varphi$ can be determined by minimizing the entanglement entropy.

\end{appendices}


\begin{thebibliography}{82}
\bibitem [{note6()}]{note6}%
	\BibitemOpen
	\bibinfo {note} {There are two kinds of notations which we have used: one is
		the vector notation $\vec{\tau}=\tau_1\vec{e}_x+\tau_2\vec{e}_y$ and the
		other is the complex number notation $\tau=\tau_1+i\tau_2$. They are
		equivalent and describe the same torus shape.}\BibitemShut {Stop}%
\end{thebibliography}
\end{document}